\documentclass[12pt,a4paper]{article}

\usepackage{epsf}
\usepackage{amsmath}
\usepackage{bbm}
\usepackage{rotate}
\usepackage{rotating}
\renewcommand{\baselinestretch}{1.5}

\newcommand{\be}{\begin{equation}}
\newcommand{\ee}{\end{equation}}
\newcommand{\bea}{\begin{eqnarray}}
\newcommand{\eea}{\end{eqnarray}}
\newcommand{\refeq}[1]{(\ref{#1})}
\newcommand{\bt}{\tilde{b}}
\newcommand{\gt}{\tilde{g}}
\newcommand{\gh}{\hat{g}}
\newcommand{\wh}{\hat{\omega}}
\newcommand{\rmd}{\mbox{\rm{d}}}
\newcommand{\ad}{\dot{\alpha}}
\newcommand{\pd}{\dot{\phi}}
\newcommand{\add}{\ddot{\alpha}}

\newcommand{\mscr}[1]{\mbox{\scriptsize #1}}
\newcommand{\ft}[2]{{\textstyle\frac{#1}{#2}}}
\begin{document}
\begin{titlepage}
\begin{center}

\hfill hep-th/0403063\\
\hfill FSU-TPI-02/04

\vskip 1cm {\large \bf  Phase Space Analysis of Quintessence Cosmologies\\[1.2ex] \bf with a Double Exponential Potential}\footnote{Work supported by the `Schwerpunktprogramm Stringtheorie' of the DFG.}

\vskip .5in

{\bf Laur J\"arv, Thomas Mohaupt and Frank Saueressig }  \\

{\em Institute of Theoretical Physics,
Friedrich-Schiller-University
Jena, \\
 Max-Wien-Platz 1, D-07743 Jena, Germany}\\
{\tt L.Jaerv, T.Mohaupt, F.Saueressig@tpi.uni-jena.de}

\vskip 0.5cm

\end{center}
\vskip 1.5cm

\begin{center} {\bf ABSTRACT } \end{center}

\noindent
We use phase space methods to investigate closed, flat, and open
Friedmann-Robertson-Walker cosmologies with a scalar potential
given by the sum of two exponential terms. The form of the potential is
motivated by the dimensional reduction of M-theory with non-trivial
four-form flux on a maximally symmetric internal space. To describe the
asymptotic features of run-away solutions we introduce
the concept of a `quasi fixed point.' We give the complete
classification of solutions according to their late-time behavior
(accelerating, decelerating, crunch) and the number of periods of
accelerated expansion.

\end{titlepage}

\tableofcontents
\setcounter{footnote}{0}
\begin{section}{Introduction}
The two most intriguing (and supposedly related) recent results in observational cosmology tell that the universe is currently undergoing an accelerated expansion while the main contribution to its energy density comes from dark energy \cite{Spergel:2003cb, Tonry:2003zg, Knop:2003iy}.
As the origin of the dark energy remains elusive from the point of view of standard cosmology and particle physics \cite{Carroll:2003qq}, a number of possible explanations have been put forth, ranging from postulating the existence of a non-zero cosmological constant or a scalar field with a suitable potential (a form of quintessence \cite{Caldwell:1997ii}) to modifications of general relativity.
Many of these ad-hoc models are in agreement with the observational data, therefore it is interesting to investigate those which can be derived from a fundamental, unifying theory of all interactions, a good candidate for such being string/M-theory.

One particular and relatively simple approach in this direction, which has received considerable attention over the last year, is string or M-theory compactified on a time-dependent maximally symmetric internal manifold \cite{Townsend:2003fx}, optionally adding the flux of a covariantly constant four-form field strength
\cite{Ohta:2003pu, Roy:2003nd, Wohlfarth:2003ni, EG}.
Given a non-flat geometry of the internal space, the resulting four-dimensional low energy effective action involves a scalar field (stemming from the dynamical volume of the internal space) governed by a potential with one exponential term. Switching on flux adds another exponential term.
This setup quite generically leads to a transient period of accelerated cosmological expansion inducing a modest increase of the scale factor.
In refs. \cite{Chen:2003ij, Ish1} it was pointed out that for certain values of parameters also eternally accelerating expansion is possible, albeit the acceleration tends to zero at late times.
In refs. \cite{Wohlfarth1, Townsend:2003qv, CC, Vieira} these features  were qualitatively inferred by relying on an earlier work due to Halliwell \cite{Halliwell} based on phase space methods.

Halliwell's original phase space analysis considered a scalar field with a single
exponential potential in a homogeneous and isotropic universe.
Subsequently it was extended to include barotropic matter \cite{Wands_etal}
and multiple scalar fields \cite{vandenHoogen:cf,Li:2001xa,Guo:2003eu,Rome},
and generalized for anisotropic universes \cite{anisotropic}.
The phase space analysis of cosmologies arising from toroidal compactifications of string and M-theory with a four-form flux (inducing a single exponential potential) was carried out in refs. \cite{Billyard:1999dg,Coley:2002xb}.
Potentials given by the sum of two
exponential terms have been used to model quintessence in refs.
\cite{Barreiro:1999zs,GAI,Majumdar:2001mm}.
The aim of the current paper is to systematically analyze the dynamics of the double exponential case using phase space methods. Even though our main interest is in potentials arising from string or M-theory compactifications, we have not limited our study to these cases and effectively cover all possible positive ratios of the coefficients in the exponentials.

Phase space methods are particularly useful when the equations of motion are hard to solve analytically.\footnote{The general analytic solution of flat FRW universe and a scalar field with a single exponential potential was recently found in ref. \cite{Russo:2004ym}, while for certain values of parameters general analytic solutions are also known when barotropic matter is added \cite{Rubano:2001su, Cardenas:2001jh, Dehnen:2002ni}.} Computing numerical solutions with random initial conditions is not a satisfying alternative as this may not reveal all the interesting properties.
To capture the features of solutions where the scalar field runs away to
infinity we introduce the concept of a `quasi fixed point': a fixed point of the reduced dynamical system which describes the asymptotic limit of run-away solutions. In this manner we gain results analogous to the single exponential case, as expected \cite{Wohlfarth1}, since in the asymptotic regime one of the exponential terms dominates over the other. New features occur in the regime where both exponential terms are important. These include an unstable de Sitter fixed point which, however, cannot be obtained from maximally symmetric compactifications. Combining the information from the fixed point analysis and numerical solutions, we are able to give the complete classification of solutions according to their late-time behavior.
In particular we discover a new class of M-theory cosmologies with two accelerating phases, one transient and another eternal.\footnote{When this manuscript was in preparation, ref. \cite{Ishproc} appeared, mentioning a possibility for two periods of acceleration in this setup, but without giving specifics.}

The paper is organized as follows.
In section 2 we begin by considering a higher-dimensional gravitational action with a flux term and dimensionally reduce it on a maximally symmetric internal space which can be either hyperbolic, toroidal, or spherical. We use the Friedmann-Robertson-Walker (FRW) ansatz with a non-trivial lapse function to obtain the equations of motion which we write in the form of a dynamical system. This helps us to map out the structure of the phase space (regions corresponding to acceleration or deceleration and to open, flat, or closed universes) and to obtain detailed information about the various fixed points.
With these analytic results at hand, we discuss the phase portraits in section 3. There we also give a classification of all possible solutions, according to their late-time behavior (accelerating, decelerating, singular) and their number of phases of accelerated expansion.
In section 4 we consider solutions at the fixed points and investigate their physical properties, {\it e.g.}, the total energy density and the state equation parameter. We also observe that in all cases the internal space undergoes decelerated expansion.
Finally in section 5 we conclude with some remarks on the viability of this setup in the light of cosmological observations and give a brief outlook on string theory cosmology.
\end{section}

\begin{section}{Analysis of the Equations of Motion}
In this section we first review the derivation of the four-dimensional effective action obtained from the dimensional reduction of $(d+m)$-dimensional gravity coupled to $m$-form flux on an $m$-dimensional maximally symmetric internal manifold (see for example \cite{Wohlfarth1}). Taking the $d$-dimensional space-time to be four-dimensional FRW we rewrite the resulting equations of motion as a dynamical system and analyze the properties of the corresponding phase space analytically.
\begin{subsection}{The higher-dimensional input}
Our starting point is a gravitational theory in $(d+m)$ dimensions coupled to an antisymmetric $m$-form field strength,
\be
S^{(d+m)} = \frac{1}{\kappa^{(d+m)}} \, \int \, \rmd^d x \, \rmd^m y \, \sqrt{-g} \, \left( R - \frac{1}{2 \, m!} \, \left( F_m \right)^2 \right) \, .
\ee
This type of ansatz also includes M- and type II string theory with the dilaton consistently set to zero. We take the flux $F_m$ to be proportional to the volume form of the internal space,\footnote{The most general flux compatible with a homogeneous and isotropic space-time has been considered in \cite{Wohlfarth1}, where it was found that all other possibilities are either dual to this ansatz or do not contribute to the scalar potential.}
\be
F_m = b \, \mbox{Vol}_{\rm int} \, .
\ee
In the case of M-theory and $d=4$, $F_m$ is a seven-form, which is dual to the four-form field strength which usually appears in the action. Notice, since this seven-form carries indices taking values in the internal space only, its Chern-Simons term vanishes, so that there is no obstruction dualizing to a four-form.

We then make the following ansatz for the dimensional reduction to $d$ dimensions:
\be\label{Metric1}
\rmd s^2 = \gt_{\mu \nu}(x) \,  \rmd x^\mu \, \rmd x^\nu + \gh_{mn}(x,y) \, \rmd y^m \, \rmd y^n \, .
\ee
Here $\gt_{\mu \nu}$ denotes the $d$-dimensional space-time metric (with coordinates $x$), and $\gh_{mn}$ is the metric on the internal space (with coordinates $y$) which takes the form
\be\label{Metric2}
\gh_{mn} = e^{2 \beta(x)} \, \wh_{mn}(y) \, .
\ee
Here $e^{2 \beta(x)}$ corresponds to a warp-factor of the internal space and $\wh_{mn}$ is the normalized metric on the maximally symmetric internal space satisfying
\be\label{Metric3}
\hat{R}_{mn}(\wh) = k_1 \, \left( m-1 \right) \, \wh_{mn}(y) \, , \quad \quad \quad \int \rmd^{m} y  \, \sqrt{- \wh(y)} = 1 \, .
\ee
The type of internal space indicated by $k_1$ can be spherical $(k_1 = +1)$, toroidal ($k_1 = 0$), or hyperbolic $(k_1 = -1)$, where the latter has been made compact by dividing out a suitable discrete isometry.

Upon dimensional reduction to $d = 4$ and by performing a conformal rescaling of the space-time metric $\gt_{\mu\nu} \rightarrow g_{\mu\nu}$, one obtains the Einstein frame effective action
\be\label{1.1}
S^{(4)} = \int \rmd^4x \sqrt{-g} \left( \frac{1}{2} R - (\partial \phi)^2 - 2 V(\phi) \right) \, ,
\ee
with the scalar potential
\be\label{1.2}
V(\phi) = - \, k_1 \, e^{-2 c \phi} + \frac{\bt^2}{2} e^{-\frac{6}{c} \phi} \, .
\ee
Here we have set the gravitational couplings to unity,
$\kappa^{(4)} = 1$, $\kappa^{(4+m)} = 1$,
and introduced $c = \sqrt{\frac{m+2}{m}}$. 
The quantities $\bt$ and $\phi$ are related to their microscopic counterparts $b$ and $\beta$ by
\be
b^2 =: 4 \, \bt^2 \, \left( \frac{4}{m \, (m-1) } \right)^{-3/c^2} \, ,
\ee
and
\be\label{1.2a}
\beta(x) = \frac{2}{m \, c} \, \phi(x) + \frac{1}{m+2} \, \ln \left( \frac{m (m-1)}{4} \right) \, ,
\ee
respectively.

By dimensional reduction we can obtain $ 1 \le c \le \sqrt{3}$ only, where the case $c = \sqrt{3}$ corresponds to compactification on $S^1$. Nevertheless we will analyze the full interval $ 0 < c < \infty$ for two reasons, first for the sake of completeness and second because other types of compactifications might lead to $c$-values outside the interval $ 1 \le c \le \sqrt{3}$.

\end{subsection}
\begin{subsection}{Four-dimensional FRW cosmology}
Let us now consider FRW space-times with a non-trivial lapse function,
\be\label{1.4}
\rmd s^2 = - N(t)^2 \rmd t^2 + e^{2 \alpha(t)} \, \left( \frac{1}{1 - k \, r^2} \rmd r^2 + r^2 \rmd \Omega_2^2
\right) \, .
\ee
Here $\rmd\Omega_2$ denotes the line element of the two-sphere and $k = -1, 0, 1$ corresponds to an open, flat, or closed universe, respectively. Using this ansatz we obtain Friedmann's equation,
\be\label{1.6}
\frac{\ad^2}{N^2} + k \, e^{-2 \alpha} = \frac{1}{3} \frac{\pd^2}{N^2} + \frac{2}{3} \, V(\phi) \, ,
\ee
as well as the dynamical equations,
\bea\label{1.7}
\frac{1}{N} \, \frac{d}{dt} \left( \frac{\ad}{N} \right) - k \, e^{-2 \alpha} + \frac{\pd^2}{N^2} & = &  0 \, , \\ \label{1.8}
\frac{1}{N} \, \frac{d}{dt} \left( \frac{\pd}{N} \right) + \frac{3}{N^2} \, \ad \pd + \frac{\partial V(\phi)}{\partial \phi} & = & 0 \, ,
\eea
where the dot denotes a derivative with respect to the time variable $t$.
The cosmological time $\tau$ is related to $t$ via $\rmd \tau = N(t) \rmd t$.

In order to study these equations as a dynamical system we proceed as follows. First we use Friedmann's equation to eliminate the term containing
$\alpha(t)$ from the dynamical equations. Then we specify the lapse function $N(t)$:
\be\label{1.9}
\begin{array}{rclll}
N(t) & := & \frac{\sqrt{2}}{\bt} \, e^{\frac{3}{c} \, \phi(t)} \, , \qquad & k_1 = 0 \, , & \\
N(t) & := & e^{c \phi(t)} \, , & k_1 = \pm 1 \, , & c < \sqrt{3} \, ,  \\
N(t) & := & \frac{\sqrt{2}}{\bt} \, e^{\frac{3}{c} \, \phi(t)} \, , & k_1 = \pm 1 \, , & \sqrt{3} < c \, .
\end{array}
\ee
%
%
For finite values of $\phi(t)$ these lapse functions are positive definite, implying this is an  admissible choice. As long as no approximations are made different lapse functions lead to different parametrizations of the same physics. The choice (\ref{1.9}) is motivated by the regime $\phi \rightarrow \infty$, where one of the exponentials appearing in the potential \refeq{1.2} becomes unimportant and may be truncated. In this approximation the effect of the leading exponential term is encoded in the definition of our time variable. As explained below eq. (\ref{4.6}) this procedure gives the correct late-time behavior of the solutions. The case $c = \sqrt{3}$, $k_1 = \pm 1$ is special, as for this particular value of $c$ the exponentials appearing in the potential are the same, $V = \left( \frac{\bt^2}{2} - k_1 \right) \, e^{-2 \sqrt{3} \phi}$. Therefore this case can be treated analogously to the single exponential case, $k_1 = 0$, and will not be considered explicitly.

Finally we introduce the variables
\be\label{1.11a}
x(t) := \phi(t) \, , \quad y(t) := \pd(t) \, , \quad z(t) := \ad(t) \, ,
\ee
in order to rewrite the dynamical equations \refeq{1.7}, \refeq{1.8}
as a set of autonomous first order differential equations.
In the single exponential case ($k_1=0$) we obtain the two-dimensional
dynamical system ${\cal D}'(y,z)$:
\be\label{1.14}
\left\{
\begin{array}{rl}
\dot{y} = & \; \frac{3}{c} \, y^2 - 3 \,z \, y + \frac{6}{c} \,  , \\
\dot{z} = & \; \frac{3}{c} \, y \, z - \frac{2}{3} y^2 - z^2 + \frac{2}{3} \, .
\end{array}
\right.
%
\ee
Note that we do not take the trivial equation $\dot{x}=y$ to be part of the dynamical system. This is possible because the equations for $y$ and $z$ do not depend on $x$, as a result of the choice of the lapse function \refeq{1.9}. Therefore we have an autonomous system which only involves $y$ and $z$. This system has been studied in refs. \cite{Halliwell,CC,Vieira}.

In the double exponential case ($k_1 = \pm 1$) we obtain the
three-dimensional dynamical system ${\cal D}(x,y,z)$:
\be\label{1.19}
\begin{array}{l}
\left\{
\begin{array}{rcl}
\dot{x} & = & y \, , \\
\dot{y} & = & c \, y^2 - 3 \, y \, z - 2 \, c \, k_1 + \frac{3 \, \bt^2}{c} e^{\frac{2}{c} \, \left( c^2 - 3 \right) x } \, ,  \\
\dot{z} & = & c \, y \, z - \frac{2}{3} \, y^2 - z^2 - \frac{2}{3} \, k_1 + \frac{\bt^2}{3} \, e^{\frac{2}{c} \, \left( c^2 - 3 \right) x } \, ,  \qquad c < \sqrt{3} \, ,
\end{array}
\right.
\\
\left\{
\begin{array}{rcl}
\dot{x} & = & y \, , \\
\dot{y} & = & \frac{3}{c} \, y^2   - 3 \, y \, z + \frac{6}{c} - \frac{4 c k_1}{\bt^2} e^{- \frac{2}{c} \, \left( c^2 - 3 \right) x} \, , \\
\dot{z} & = & \frac{3}{c} \, y \, z - \frac{2}{3} \, y^2 - z^2 + \frac{2}{3} - \frac{4 k_1}{3 \bt^2} \, e^{- \frac{2}{c} \left( c^2 - 3 \right) x} \, ,  \qquad \sqrt{3} < c \, .
\end{array}
\right.
\end{array}
\ee
%
%
In this case we cannot drop the equation $\dot{x}=y$, because the
equations for $y$ and $z$ depend on $x$. The $x$-dependent terms
correspond to the subleading exponential term, while the leading
one, which dominates for large $\phi$,
was absorbed by the choice \refeq{1.9} of the lapse function.
\end{subsection}
\begin{subsection}{The structure of the phase space}
After defining the variables parametrizing the phase spaces of our systems, we will now discuss their structure. As the dimension of the phase space for $k_1 = 0$ and $k_1 = \pm 1$ is different, we treat them separately in the following.

We start by investigating which points in phase space are associated with curvature singularities. To this end we compute the Ricci-scalar of the metric \refeq{1.4},
\be\label{1.5}
R = \frac{6}{N^2} \left( \add + 2 \ad^2 - \frac{\dot{N}}{N} \ad + k N^2 e^{-2 \alpha} \right) \, .
\ee
Substituting in Friedmann's equation, the specific choice of the lapse function \refeq{1.9}, and the dynamical equation \refeq{1.7} this expression takes the following form:
\be\label{1.10}
\begin{array}{rclll}
R & = & \bt^2 \, e^{- \, \frac{6}{c} \, x} \, \left( 4 - y^2 \right) \, , & k_1 = 0 \, , &  \\
R & = & 2 \, e^{-2 c x} \left( -4 k_1 + 2 \, \bt^2 \, e^{\frac{2}{c} \, \left( c^2 - 3 \right) \, x} - y^2 \right) \, , \quad & k_1 = \pm 1 \, , & c < \sqrt{3} \, , \\
R & = & \bt^2 \, e^{- \, \frac{6}{c} x} \, \left( 4 - \frac{8 k_1}{\bt^2} \, e^{- \frac{2}{c} \left( c^2 - 3 \right) x} - y^2  \right) \, , & k_1 = \pm 1 \, , & \sqrt{3} < c \, .
\end{array}
\ee
%
%
These equations indicate that for both $k_1 = 0$ and $k_1 = \pm 1$ the boundaries of the phase space $x = - \infty$ and $y = \pm \infty$ correspond to curvature singularities, while at the surfaces $z = \pm \infty$ as well as $x = + \infty$ the curvature scalar remains finite. This result also holds for other curvature invariants as $R_{\mu \nu} \, R^{\mu \nu}$ and $R_{\mu \nu \rho \sigma} \, R^{\mu \nu \rho \sigma}$.

Next we determine the phase space regions, where the solutions undergo acceleration in the scale factor
%
$a(\tau)=e^{\alpha(t(\tau))}$. These are characterized by the inequality
\be\label{1.11}
a(\tau)^{\prime\prime} := \frac{d^2}{d\tau^2} a(\tau) = \left( \frac{1}{N} \, \frac{d}{dt} \right)^2 e^{\alpha(t)} > 0 \, .
\ee

Substituting in the lapse functions \refeq{1.9} together with the dynamical equation \refeq{1.7}, this implies that acceleration occurs for
\be\label{1.12}
\begin{array}{rclll}
y^2 & < & 1 \,  , & k_1 = 0 \, , & \\
y^2 & < & - k_1 + \frac{ \bt^2 }{2} \, e^{\frac{2}{c} \, \left( c^2 - 3 \right) \, x}  \, , \quad & k_1 = \pm 1 \, , & c < \sqrt{3} \, , \\
y^2 & < & 1 - \frac{2 k_1}{\bt^2} \, e^{-\frac{2}{c} \left( c^2 - 3 \right) x} \, , & k_1 = \pm 1 \, , & \sqrt{3} < c \, .
\end{array}
\ee
Likewise the regions of expansion are given by $ \frac{1}{N} \, \frac{d}{dt} e^{\alpha(t)} > 0 $, which corresponds to $z > 0$.

From eq. \refeq{1.12} we observe that acceleration occurs if $\pd^2$ (and hence the kinetic energy of the scalar field) is small. We further find that the flux contribution enlarges the acceleration region if $c < \sqrt{3}$ or $k_1 = -1$ while for  $\sqrt{3} < c$, $k_1 = +1$ the flux narrows the acceleration region.

Finally, we can use Friedmann's equation to separate the phase space into regions corresponding to closed, flat, or open universes. Setting $k=0$ and substituting in eq. \refeq{1.9} we obtain the $k=0$ hypersurfaces:
\be\label{1.13}
\begin{array}{ll}
y^2 - 3 \, z^2 + 2 = 0 ,  & k_1 = 0 \, , \\
y^2 - 3 \, z^2 - 2 \, k_1 + \bt^2 \, e^{ \frac{2}{c} \, \left( c^2 - 3 \right) \, x} = 0 \, , \quad & k_1 = \pm 1 \, , \,  c < \sqrt{3} \, , \\
y^2 - 3 \, z^2 + 2 - \frac{4 \, k_1}{\bt^2} \, e^{- \frac{2}{c} \left( c^2 - 3 \right) x} = 0 \, , & k_1 = \pm 1 \, , \,  \sqrt{3} < c \, .
\end{array}
\ee
For the left hand side being smaller (bigger) then zero, we are in the $k = -1$ ($k = +1$) region, respectively.
\end{subsection}

\begin{subsection}{Fixed points}
Let us now determine the
fixed points (critical points, equilibrium points) of the dynamical systems
${\cal D}'(y,z)$ \refeq{1.14}
and ${\cal D}(x,y,z)$ \refeq{1.19}. Recall that for
a single exponential potential ($k_1=0$) it is possible to
eliminate both $\alpha$ (using Friedmann's equation) and
$x=\phi$ (using the lapse function \refeq{1.9}), so that
the dynamical system is two-dimensional.
The fixed points (FPs) are given by
\be\label{1.14a}
\dot{y} \left. \right|_{\rm FP} = 0 \, , \quad \dot{z} \left. \right|_{\rm FP} = 0 \, .
\ee
At a FP the scalar field $\phi(t)$ and the scale
factor $\alpha(t)$
therefore have the following form:
\be\label{1.15a}
\phi(t) = \pd^* \, t + c_1 \, , \quad \alpha(t) = \ad^* \, t + c_2 \, .
\ee
Note that $\phi$ does not go to a constant at the FPs
of ${\cal D}'(y,z)$. That would be the case if we looked
for FPs of the larger system ${\cal D}(x,y,z)$, which is
obtained by adding the third equation $\dot{x}= y$ to \refeq{1.14}.
However, for a single exponential potential ${\cal D}(x,y,z)$ does not
have FPs. This is clear, because the potential does not
have an extremum. The generic late-time behavior of
eternal solutions is a run-away behavior $\phi \rightarrow \infty$
with asymptotically constant velocity (with respect to the
time variable $t$), corresponding to the FPs of
${\cal D}'(y,z)$.

For of the double exponential potential it is not possible
to eliminate $x=\phi$ from the equations for $y$ and $z$
and we have to deal with the three-dimensional dynamical system
${\cal D}(x,y,z)$ \refeq{1.19}. The FPs of this system satisfy
the equations
\be\label{1.25}
\dot{y} \left. \right|_{\rm FP} = 0 \, , \quad \dot{z} \left. \right|_{\rm FP} = 0  \; \mbox{and} \; \dot{x} \left. \right|_{\rm FP} = 0 \, ,
\ee
and solutions for $\phi(t)$ and $\alpha(t)$ at the FP take the form
\be
\phi(t) = \phi^* \, , \quad \alpha(t) = \ad^* \, t + c_1 \, .
\label{FPxyz}
\ee
A double exponential can in principle have
an extremum at a finite value of $\phi$, and we will see later that for a certain
range of parameters the corresponding FP indeed
exists.

We also expect that solutions with the run-away
behavior \refeq{1.15a} exist and play a role in the dynamics of
${\cal D}(x,y,z)$. Observe that for $\phi \rightarrow  \infty$
the $x$-dependent terms in \refeq{1.19} become exponentially small.
Therefore it is a reasonable approximation to drop them,
which leads to a reduced two-dimensional dynamical system
${\cal D}_{\infty}(y,z)$.\footnote{If one just drops the first
equation in \refeq{1.19} the resulting non-autonomous system
does not have FPs for finite $x$.}
For $\sqrt{3}<c$ this
is in fact the same system as the one obtained for $k_1=0$,
${\cal D}'(y,z)$ \refeq{1.14}. The physical picture behind
this truncation is that solutions which move towards
$\phi=\infty$ enter an intermediate regime of large but finite
$\phi$, where the subleading
exponential becomes irrelevant and can be dropped, while the
leading exponential is encoded in the lapse function \refeq{1.9}.
The behavior in this regime is then controlled by the
reduced dynamical system ${\cal D}_{\infty}(y,z)$, {\em i.e.},
cosmological solutions adjust to
the FPs of ${\cal D}_{\infty}(y,z)$ (or are repelled from,
depending on the nature of the FP). In the
three-dimensional parameter space $(x,y,z)$ we expect that
the solutions flow in a narrow tube towards
the attractive FPs of ${\cal D}_{\infty}(y,z)$, which are located at
the boundary $x=\infty$.
The FPs
of ${\cal D}_{\infty}(y,z)$ will be referred to as
quasi fixed points (QFPs) of the full system ${\cal D}(x,y,z)$
in the following. At such QFPs, the time-dependence
of $\phi(t)$ and $\alpha(t)$ is the same as the one found for
$k_1 = 0$, \refeq{1.15a}. For terminological convenience we will also refer to the FPs of the reduced system ${\cal D}'(y,z)$ as QFPs.\footnote{Note that the FPs discussed in \cite{Halliwell, CC, Vieira} are QFPs in our sense of definition.}

We remark that the above definition of a QFP is
tied to a particular,
physically motivated approximation, namely dropping the sub-leading
exponential.
The effect of the leading exponential is captured by
the lapse function, which therefore acquires a physical meaning.
Comparing to numerical solutions of the full equations of motion
in the next section, we establish that the
picture outlined  above is indeed correct and that the concept of QFPs
is useful for describing the dynamics of
${\cal D}(x,y,z)$.  Also note that the behavior of $\phi(t)$ and
$\alpha(t)$ at a QFP of ${\cal D}(x,y,z)$
involves a distinguished time coordinate. In contrast to a
true FP in phase space, \refeq{1.15a} is not reparametrization
invariant. Of course, reparametrization invariance as such is not an
issue, since we use a specific (class of) parametrization(s),
\refeq{1.4}, anyway. However, the time variable $t$ does not have
a direct physical interpretation. Therefore we will
investigate the analytical properties of  (Q)FP solutions
with respect to the
cosmological time $\tau$ in section 4.

Whether a FP or QFP is an attractor, in the sense that there is a class of solutions which asymptotically evolve with the same values $\pd^*, \ad^*$ (and $\phi^*$), can be deduced form the eigenvalues of the stability matrix
\be\label{1.17}
\left. {\bf B}_{ij} \right|_{\rm FP} = \left. \partial_j \beta_i \, \right|_{\rm FP} \, .
\ee
Here $\beta_i$ is short for the right hand sides of eqs. \refeq{1.14} and \refeq{1.19} and $i \in \{ y, z \}$, $\vec{X} = \left[y,z\right]^{\rm T}$,  or $i \in \{x, y, z \}$, $\vec{X} = \left[x,y,z\right]^{\rm T}$, depending on whether we evaluate the condition \refeq{1.14a} or \refeq{1.25}. Denoting the eigenvectors and eigenvalues of this matrix as $\vec{v}_\alpha$ and $\theta_\alpha$, respectively, we can write down the linearized solution of the full system in the vicinity of a (hyperbolic) FP,
\be\label{1.17a}
\vec{X}(t) = \vec{X}^* + \sum_\alpha \, c_\alpha \, \vec{v}_\alpha \, e^{\theta_\alpha \, t}  \, ,
\ee
with $c_\alpha$ being the constants of integration. This indicates an attractor has $\theta_{\alpha} < 0$ for all eigenvalues, while for mixed positive and negative eigenvalues, one obtains a saddle point.

An interesting subtlety in the attractor behavior arises if the
(Q)FP is a saddle point located on the $k=0$ surface. In this case
it is necessary to check the direction of the eigenvectors
relative to the normal of the surface to deduce for which
solutions the (Q)FP acts as an attractor. For $z>0$ (expansion
region) the typical configuration arising here is that the
attractive eigendirections are tangent to the surface while the
repulsive ones have a non-zero component along the normal vector,
such that the (Q)FP acts as an attractor for $k = 0$ solutions
while solutions starting in the regions $k = \pm 1$ are repelled
and cannot reach it. For $z<0$ the opposite situation occurs.
\subsubsection*{The case $k_1 = 0$}
We start by analyzing the QFPs of eq. \refeq{1.14}. Explicitly evaluating the QFP condition \refeq{1.14a} yields that the system has two pairs of QFPs:\footnote{Due to the invariance of \refeq{1.14} under time-reversal, (Q)FPs always come in pairs, {\it i.e.}, if $[\pd^*, \ad^*]$ is a (Q)FP, so is $[-\pd^*, -\ad^*]$. Further the eigenvalues of the second ones are given by the negative of the eigenvalues of the first, so that in the following we mention the first ones only.}
\be\label{1.15}
\begin{array}{lll}
\mbox{QFP}_1 = & \left[ \pm 1 \, , \, \pm \frac{3}{c} \right] \, , & \mbox{exists for} \; 0 < c < \infty \, , \\
\mbox{QFP}_2 = & \left[ \pm \sqrt{\frac{6}{c^2 - 3}} \, , \, \pm \frac{c}{3} \, \sqrt{\frac{6}{c^2 - 3}} \,  \right] \, , & \mbox{exists for} \;  \sqrt{3} < c \, .
\end{array}
\ee
Substituting the QFP coordinates into the equations for the acceleration region \refeq{1.12} and the $k=0$ surface \refeq{1.13} yields that  $\mbox{QFP}_1$ is located in the $k = -1$ and $k = +1$ region for $0<c<3$ and $3 < c$, respectively. Further it is always situated at the boundary between the acceleration and deceleration region. The $\mbox{QFP}_2$ always resides on the $k=0$ surface, being in the deceleration region for $ \sqrt{3} < c < 3$ and in the acceleration region for $3 < c$. The case $c = 3$ is special. Here the two QFPs merge into a single non-hyperbolic QFP at $[ \pm 1 , \pm 1 ]$.

To determine the properties of the phase space trajectories close to the QFPs, we calculate the eigenvectors and eigenvalues of the stability matrix \refeq{1.17}:
\be\label{1.18}
\begin{array}{ll}
\theta_{1 +}  =   \frac{1}{c} \left( -3 + \sqrt{4 c^2 - 27} \right) \, , & \quad \theta_{1 -}   =   \frac{1}{c} \left( -3 - \sqrt{4 c^2 - 27} \right) \, ,   \\[1.5ex]
\theta_{2 + }  = \frac{1}{c} \sqrt{\frac{6}{c^2 - 3}} \, \left( 6 - \frac{2}{3} \, c^2 \right) \, , & \quad
\theta_{2 -}  =  \, \frac{1}{c} \sqrt{\frac{6}{c^2 - 3}} \, \left( 3 -  c^2 \right) \, .
\end{array}
\ee
These eigenvalues imply that the $\mbox{QFP}_1$ is a complex attractor ({\it i.e.}, the solutions spiral into this point), a real attractor, and a saddle point for $0 < c < \frac{3}{2} \sqrt{3}$,  $\frac{3}{2} \sqrt{3} \le c < 3$ and $3 < c $, respectively. Analogously one obtains that the $\mbox{QFP}_2$  is a saddle point for $\sqrt{3} < c < 3$ and becomes an attractor for $3<c$. These results are also summarized in table \ref{t.1}. Note that this table includes the QFPs in the expansion region ($z>0$) only.

In the regime $\sqrt{3} < c < 3$ the QFP$_2$ is a saddle point on the $k=0$ surface. By comparing the eigendirections of the QFP with the normal of the surface we find that the attractive eigendirection is tangent to the $k=0$ surface, while the repulsive eigendirection has a non-trivial component along the normal. This indicates that the $\mbox{QFP}_2$ acts as an attractor for solutions on the $k=0$ surface while trajectories starting in the $k = \pm 1$ regions are repelled and cannot approach it asymptotically. These features are also illustrated in the second diagram of figure \ref{eins}.
\begin{table}[tp!]
\renewcommand{\baselinestretch}{1.3} \large \normalsize
\begin{tabular*}{\textwidth}{@{\extracolsep{\fill}} cccccc} \hline \hline
(Q)FP & $k_1$ & $c$ & type &  $k$ & acceleration \\ \hline
1  & $ 0 $ & $ 0 < c < \frac{3}{2} \, \sqrt{3} $ & complex attractor  & $  -1 $ & boundary \\
   &            & $ \frac{3}{2} \, \sqrt{3} \le c < 3 $ & real attractor  & $  -1 $ & boundary \\
   &            & $ 3 < c $ & saddle point &  $ +1 $ & boundary \\[1.5ex] \hline
2  & $ 0 $ & $ 0 < c \le \sqrt{3} $ & $-$ & $-$ & $-$ \\
   &            & $ \sqrt{3} < c < 3 $ & saddle point  & $  0 $ & deceleration \\
   &            & $ 3 < c    $ & real attractor & $  0 $ & acceleration \\
\hline
3  & $ +1 $ & $ 0 < c \le  \sqrt{3} $ & $-$  &  $-$  & $-$ \\
   &            & $ \sqrt{3} < c $ & saddle point  & $  0 $ & acceleration \\[1.5ex] \hline
4  & $ -1 $ & $ 0 < c < 1 $ & saddle point  & $ +1 $ & boundary \\
   &             & $ 1 < c \le \frac{2}{\sqrt{3}} $ & real attractor  & $ -1 $ & boundary \\
   &             & $ \frac{2}{\sqrt{3}} < c < \sqrt{3} $ & complex attractor &  $ -1 $ & boundary \\
  &              & $ \sqrt{3} \le c    $ & $ - $ & $  - $ & $ - $  \\[1.5ex] \hline
5  & $ -1 $ & $ 0 < c < 1 $ & real attractor & $ 0$ & acceleration \\
   &             & $ 1 < c < \sqrt{3}$ & saddle point  & $ 0 $ & deceleration \\
   &             & $ \sqrt{3} \le c    $ & $-$  & $ - $ & $ - $ \\
\hline
6  & $ \pm 1 $ & $ 0 < c \le \sqrt{3} $ & $ - $  & $ - $ & $ - $ \\
   &                & $ \sqrt{3} < c < \frac{3}{2} \sqrt{3} $ & complex attractor & $ -1$  & boundary \\
   &            & $ \frac{3}{2} \sqrt{3} \le c < 3 $ & real attractor  & $ -1 $ & boundary \\
   &            & $ 3 < c $ & saddle point &  $  +1 $ & boundary \\[1.5ex] \hline
7  & $ \pm 1 $ & $ 0 < c \le \sqrt{3} $ & $-$ & $-$ & $-$ \\
   &            & $ \sqrt{3} < c < 3 $ & saddle point  & $  0 $ & deceleration \\
   &            & $ 3 < c    $ & real attractor & $ 0 $ & acceleration \\
\hline \hline
\end{tabular*}
\renewcommand{\baselinestretch}{1}
\parbox[c]{\textwidth}{  \caption{\label{t.1}{\footnotesize Summary of the fixed point properties found in subsection 2.4. Here `(Q)FP' gives the number of the (quasi) fixed point as it appears in the text, while `$k_1$' and `$c$' are the parameters of the potential \refeq{1.2}. The columns `type', `$k$' and `acceleration' specify the properties of the (Q)FP for the corresponding values of the parameters. At the values of $c$ not covered in this table we encounter non-hyperbolic FPs.}}}
\end{table}
%
%
\subsubsection*{The case $k_1 = \pm 1$ }
Let us now consider the case $k_1 = \pm 1$. We start by analyzing the condition for `true' FPs. Imposing the conditions \refeq{1.25}, the only non-trivial solution requires $k_1 = +1$  and is given by
\be\label{1.20}
\mbox{FP}_3 = \left[ \frac{c}{2 \, (c^2 -3) } \, \ln \left( \frac{2 c^2}{3 \bt^2} \right) \, , \, 0 \, , \, \pm \sqrt{ \frac{2}{3 \, c^2} \, ( c^2 - 3 ) }  \right] \, , \quad \mbox{exists for} \; c > \sqrt{3} \, .
 \ee
It is located on the $k=0$ surface and inside the acceleration region. In fact, it will turn out in section 4 that this FP gives rise to an exponentially growing scale factor in cosmological time and therefore is a true de Sitter FP. Taking $i \in \{ x, y, z \}$ and calculating the eigenvalues of the stability matrix \refeq{1.17}  for this FP yields
\be\label{1.21}
\theta_3^1 = \sqrt{2 (c^2 - 3)} \, , \; \;
\theta_3^2 = - \, 2 \, \sqrt{2 (c^2 - 3)} \, , \; \;
\theta_3^3 = - \, \frac{2}{3} \, \sqrt{2 (c^2 - 3)} \, .
\ee
Hence it is a saddle point having two late-time attractive and one late-time repulsive eigendirection. Comparing its eigenvectors to the normal of the $k=0$ surface yields that one attractive and the repulsive eigendirection are tangent to the surface while the second attractive eigenvector has a non-zero component along the normal. This implies that the FP${}_3$ is a saddle point on the $k=0$ surface, {\it i.e.}, flat cosmological solutions need to be fine-tuned in order to approach it asymptotically. Additionally, we can also fine-tune solutions starting in the $k = \pm 1$ region such that they asymptotically approach this FP. The latter properties are reflected in table \ref{t.4}, while the results on the position of the FP are compiled in table \ref{t.1}.

Notice that the FP${}_3$ is located at the (positive) maximum of the
potential (\ref{1.2}) with $k_1 = +1$. This explains why it is unstable (saddle) and
corresponds to an expanding de Sitter universe. For $c<\sqrt{3}$ the
potential has a (negative) minimum, but there is no
related FP, since this situation corresponds to an anti-de Sitter
universe which contracts to a crunch represented by the boundary of the phase space.
For $k_1 = -1$ the potential has no extrema and we do not obtain any FPs in this case.

Having found the true FP of the system \refeq{1.19} we now relax the FP condition and look for QFPs satisfying \refeq{1.14a}. As previously observed, the QFPs are always located at the $x = \infty$ boundary, but fix the values $\pd^*, \ad^*$ along which the boundary is approached.

In the case $c < \sqrt{3}$, where the $k_1$ term is dominant, substituting the ansatz \refeq{1.15a} into \refeq{1.19} and dropping the subdominant exponential term only gives rise to QFPs for $k_1 = -1$:
\be\label{1.22}
\begin{array}{ll}
\mbox{QFP}_4  =  \left[ \pm 1 \, , \, \pm c \right] \, , & \; \mbox{exists for} \; c < \sqrt{3} \, , \\
\mbox{QFP}_5  =  \left[ \pm c \, \sqrt{\frac{2}{3 - c^2}} \, , \, \pm \sqrt{\frac{2}{3 - c^2}} \right] \, ,  & \; \mbox{exists for} \; c < \sqrt{3} \, .
\end{array}
\ee
The corresponding eigenvalues of the stability matrix are given by
\be\label{1.24}
\begin{array}{ll}
\theta_{4+}  = -c + \sqrt{4 - 3 \, c^2} \, , & \quad \theta_{4-}  = -c - \sqrt{4 - 3 \, c^2} \, , \\
\theta_{5+}  = \frac{4}{\sqrt{2 (3 - c^2)}} \, (- 1 + c^2) \, , & \quad \theta_{5-} = - \sqrt{2 (3 - c^2)} \, .
\end{array}
\ee
The properties of these QFPs with respect to acceleration and the $k = 0$ surface are, again, obtained by comparing the position of the QFP relative to the surfaces \refeq{1.12} and \refeq{1.13}, see table \ref{t.1}.

In the case $c > \sqrt{3}$ the flux-term gives the dominant contribution to the potential and substituting $ x = \infty$ into \refeq{1.19} is equivalent to setting $k_1 = 0$. As a consequence, we obtain the QFPs \refeq{1.15} together with the properties \refeq{1.18}. For completeness these QFPs are labeled ${\rm QFP}_6$ and ${\rm QFP}_7$, respectively, and their properties are also included in table \ref{t.1}.
\end{subsection}
\end{section}
\begin{section}{The Phase Portraits}
The greatest advantage of the phase space method is that one
can plot phase portraits, which makes the classification of
solutions easy and intuitive. Here one uses two well-known
properties of first order ordinary differential equations:
(i) trajectories (integral lines) do not cross, and
(ii) for hyperbolic FPs the local behavior of the
exact solution is uniquely determined by the linearized system.
The first point can often be used to show that the phase space
decomposes into disjoint subspaces. The second point helps the
analysis, because the hyperbolic QFPs found in the last section
govern the behavior of the trajectories in their vicinity and
lead to an universal asymptotic behavior.
 We will first explain the two-dimensional
phase spaces occurring in toroidal compactifications with flux.
These results are already covered by the literature \cite{Halliwell,CC,Vieira}, but
serve well for explaining the method. Then we turn to the
more complicated phase spaces of curved flux compactifications.
\begin{subsection}{Toroidal internal space with flux}

The properties of the two-dimensional phase spaces arising in
models with $k_1=0$  are summarized in figure \ref{eins} and table \ref{t.3}.
They include all flux compactifications with toroidal internal
space. The QFPs of this system were classified in the last
section and their properties are summarized in table \ref{t.1}. Based on this classification, there are four
different cases for the parameter $c$. Note that the classification of hyperbolic compactifications without flux is also covered by table \ref{t.3} and requires a redefinition $c \rightarrow \frac{3}{c}$.

\begin{figure}[tp!]
\renewcommand{\baselinestretch}{1}
\begin{center}
\leavevmode
\epsfxsize=0.45\textwidth
\epsffile{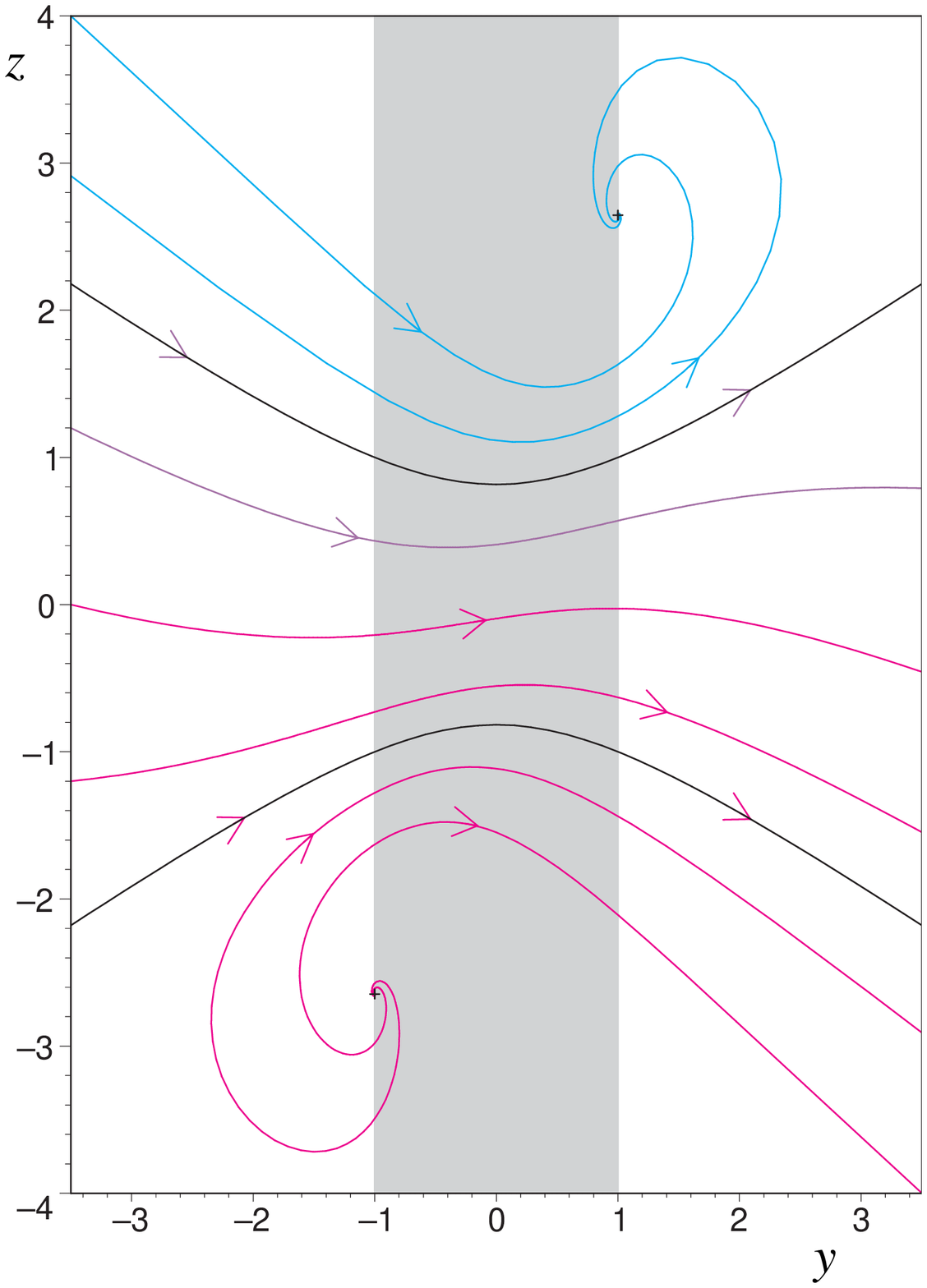} \, \, \, \, \,
\epsfxsize=0.45\textwidth
\epsffile{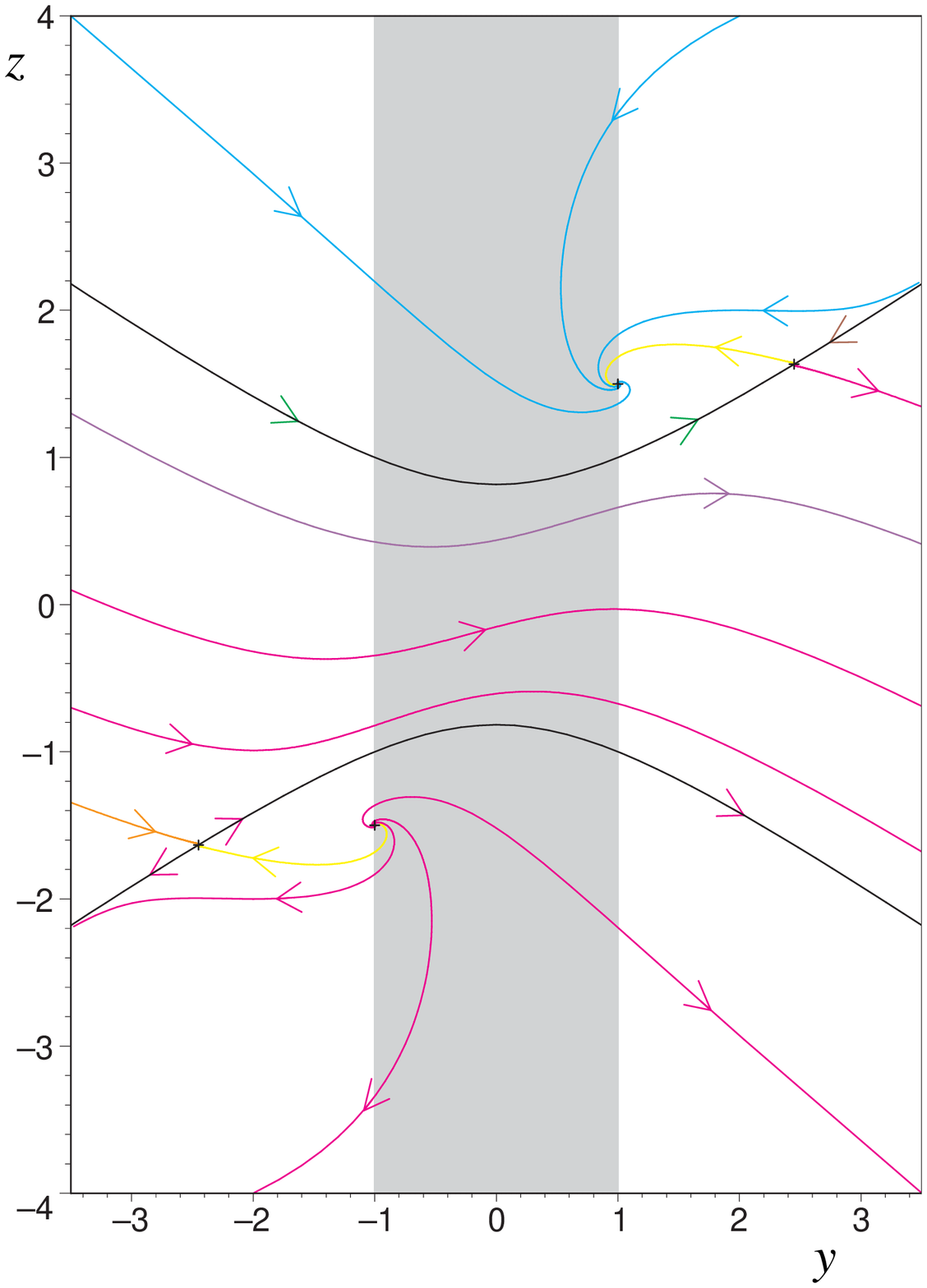} \\
%
%
\leavevmode
\epsfxsize=0.45\textwidth
\epsffile{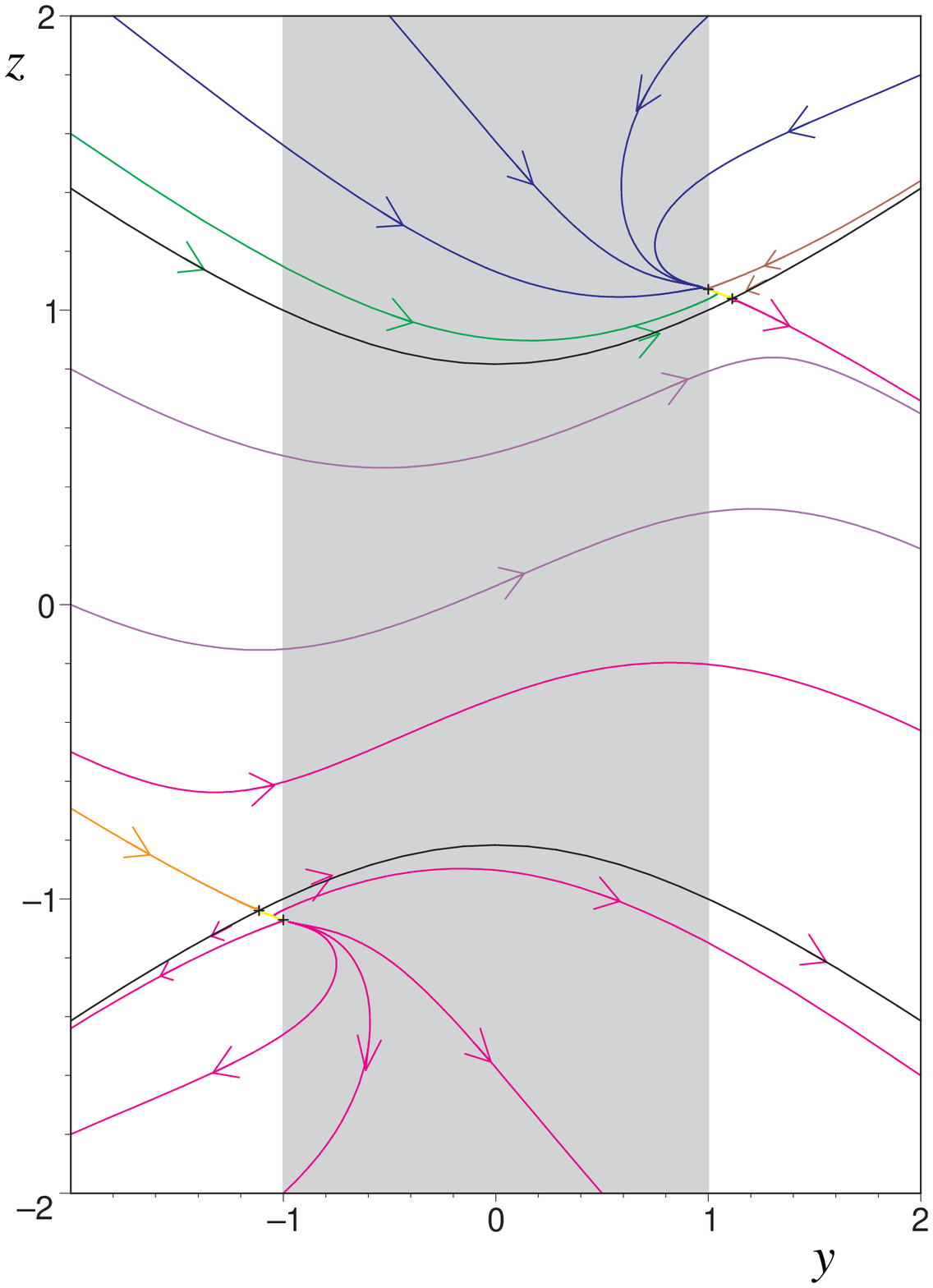} \, \, \, \, \,
\epsfxsize=0.45\textwidth
\epsffile{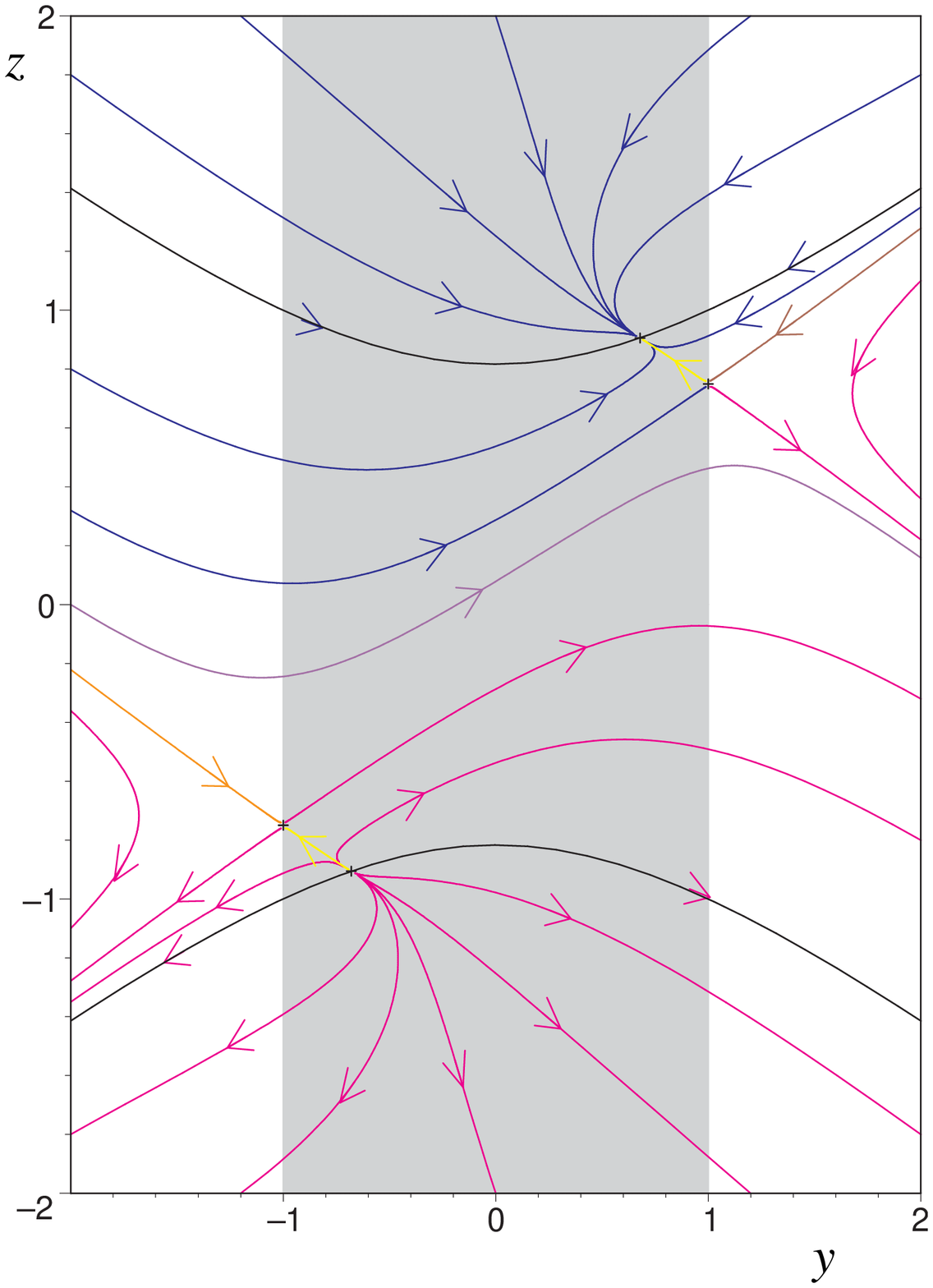}
\end{center}
\parbox[c]{\textwidth}{\caption{\label{eins}{\footnotesize Phase portraits for $k_{1} = 0$. The flow is shown for the four representative cases $c = \frac{3}{\sqrt{7}}$ (upper left), $c = 2$ (upper right), $c = 2.8$ (lower left) and $c = 4$ (lower right), respectively. The thick black lines represent the $k=0$ surface and the trajectory corresponding to a flat universe. The arrows point into the direction of increasing time variable $t$.}}}
\end{figure}

Figure \ref{eins}  shows a representative phase portrait for each of the four cases. Here
the horizontal and vertical axis correspond to $y= \dot{\phi}$
and $z=\dot{\alpha}$, respectively. The universe expands for $z>0$
and contracts for $z<0$. These two regions are
mirror images of one another as they are related by time reversal, which acts on ${\cal D}(x,y,z)$ by $x \rightarrow x$, $y \rightarrow -y$ and $z \rightarrow -z$. 
The vertical gray strip specifies the
region of acceleration determined by eq. (\ref{1.12}). More precisely,
 this region corresponds to accelerated expansion for $z>0$ and decelerated contraction for $z<0$.
Eq. (\ref{1.13}) implies that, independent of the value of $c$,
the trajectories corresponding to flat universes
(black lines) cut the phase space into three horizontal regions,
corresponding to open universes (expanding in the upper and
contracting in the lower region) and
closed universes (middle region).
%
%
The specific late-time behavior of a trajectory is encoded in its
color and can be inferred from table \ref{c.1}.
%
%
%
\begin{table}[tp!]
\renewcommand{\baselinestretch}{1.3} \large \normalsize
\begin{tabular*}{\textwidth}{@{\extracolsep{\fill}} c|l|c|l} \hline \hline
fig. & color & acc. expan. & late-time behavior \\ \hline
1 & cyan & $\infty$ & cyclic acc./dec. expansion \\
2 & red & 2 & accelerated expansion \\
1,2 & blue & 1 & accelerated expansion \\
1,2 & green & 1 & decelerated expansion \\
1,2 & dark brown & 0 & decelerated expansion \\
1 & light brown & 0 & accelerated contraction \\
1 & violet & 1 & singularity \\
1 & magenta & 0 & singularity \\
1 & yellow & depends & cross-over, starts and ends at a QFP \\
\hline \hline
\end{tabular*}
\renewcommand{\baselinestretch}{1}
\parbox[c]{\textwidth}{  \caption{\label{c.1}{\footnotesize
Colors used in figures \ref{eins} and \ref{zwei} to distinguish cosmological solutions according to their number of phases of accelerated expansion and late-time behavior.}}}
\end{table}

As an example, let us consider the phase portrait in the upper
left corner of figure \ref{eins}.
%
It corresponds to the M-theory value $c=\frac{3}{\sqrt{7}}$. In
this range of $c$ only QFP${}_1$ exists, coming in two copies:
complex attractor for $z>0$ and complex repeller for $z<0$. These
are located in the regions of open universes and at the boundary
between acceleration and deceleration. All open universes in the
upper ($z>0$) region start at a boundary and spiral into the QFP
for late times (cyan trajectories). Correspondingly, in the lower
($z>0$) region the open universes start at the QFP and spiral out
to the boundary (magenta trajectories). Flat and closed universes
run from boundary to boundary thereby
crossing the acceleration region once. Since the expansion and
contraction parts of the phase space for closed universes are not
separated, the behavior of solutions can change along
the trajectory: in the acceleration region contraction can become
expansion while in the deceleration region expansion can switch
into contraction, but not {\it vice versa}. 
Hence the closed universes can have either
one (violet) or no (magenta) phases of accelerated expansion. In summary we get two
types of late-time behavior possible in the whole setup: expanding
open universes exhibit eternal oscillations between acceleration
and deceleration, while contracting open universes as well as all
flat and closed universes terminate at a boundary of the phase
space associated with a curvature singularity.\footnote{Numerical
investigations of the behavior of the logarithmic scale factor for
closed universes indicate that the boundary is reached in finite
cosmological time and corresponds to $\alpha \rightarrow -\infty$,
$i.e.$, a big crunch.}

Similar results can be read off for the other ranges of $c$
displayed in figure \ref{eins}.
The main addition is that we do also encounter examples of
special, non-generic solutions. First, there are fine-tuned
solutions which run into or start from a saddle point and 
are thus future or past infinite, respectively. Second, there are
so-called cross-over solutions which run from one fixed point into
another and thus have both infinite past and infinite future.

\begin{table}[tp!]
\renewcommand{\baselinestretch}{1.2}
\begin{tabular*}{\textwidth}{@{\extracolsep{\fill}} ccccccccccc} \hline \hline
    &    & \multicolumn{2}{c}{} & \multicolumn{2}{c}{acc.} & \multicolumn{2}{c}{dec.} & acc. & dec. &  \\
$c$     &  $k$   & \multicolumn{2}{c}{singular} & \multicolumn{2}{c}{expan.} & \multicolumn{2}{c}{expan.} & contr. & contr. & cyclic \\
        &        &   0   &   1  & 1 & 2 &  0 & 1 & 0 & 0 & $\infty$ \\ \hline
$ 0 < c \le \sqrt{3} $
            & $-1$ &  G  & $-$ & $-$ & $-$ & $-$ & $-$ & $-$ & $-$ &  G  \\
            & $ 0$ &  G  &  G  & $-$ & $-$ & $-$ & $-$ & $-$ & $-$ & $-$ \\
            & $+1$ &  G  &  G  & $-$ & $-$ & $-$ & $-$ & $-$ & $-$ & $-$ \\ \hline
$ \sqrt{3} < c < \frac{3}{2} \sqrt{3} $
            & $-1$ &  G  & $-$ & $-$ & $-$ & $-$ & $-$ &  C  & $-$ & G,C \\
            & $ 0$ &  G  & $-$ & $-$ & $-$ &  G  &  G  & $-$ & $-$ & $-$ \\
            & $+1$ & G,F &  G  & $-$ & $-$ & $-$ & $-$ &  F  & $-$ & $-$ \\ \hline  %
$\frac{3}{2} \sqrt{3} \le c < 3 $
            & $-1$ &  G  & $-$ &  G  & $-$ & G,C &  G  &  C  & $-$ & $-$ \\
            & $ 0$ &  G  & $-$ & $-$ & $-$ &  G  &  G  & $-$ & $-$ & $-$ \\
            & $+1$ & G,F &  G  & $-$ & $-$ & $-$ & $-$ &  F  & $-$ & $-$ \\ \hline
$ 3 < c $
            & $-1$ &  G  & $-$ &  G  & $-$ & $-$ & $-$ & $-$ & $-$ & $-$ \\
            & $ 0$ &  G  & $-$ &  G  & $-$ & $-$ & $-$ & $-$ & $-$ & $-$ \\
            & $+1$ & G,F &  G  & G,C & $-$ &  F  & $-$ &  F  &  C  & $-$ \\ \hline \hline
\end{tabular*}
\renewcommand{\baselinestretch}{1}
\parbox[c]{\textwidth}{\caption{\label{t.3}{\footnotesize Classification of possible cosmological solutions for $k_1 = 0$ according to their late-time behavior and their number of periods of accelerated expansion. Here `G' denotes that the corresponding behavior is generic, `F' indicates that the behavior requires a fine-tuning of the initial conditions, and `C' marks a special solution which runs from one fixed point into another. }}}
\end{table}
The classification of all possible solutions is given in table
\ref{t.3}, which is organized as follows. The first and second
column give the values of $c$ and $k$. The following columns
characterize the late-time behavior: singular, meaning that the
solution runs to the boundary of the phase space, and accelerated
expansion ($a^{\prime\prime}>0$, $a^{\prime}>0$), decelerated
expansion ($a^{\prime\prime}<0$, $a^{\prime}>0$), accelerated
contraction ($a^{\prime\prime}<0$, $a^{\prime}<0$), decelerated
contraction ($a^{\prime\prime}>0$, $a^{\prime}<0$), or cyclic
(oscillation between accelerated and decelerated expansion) for
solutions asymptotically approaching a QFP. Since some solutions
also exhibit transient accelerated expansion, we have split the
columns further to indicate the total number of phases of
accelerated expansion. (For the cyclic case this number is infinite). 
Within the table a dash indicates that
the corresponding behavior does not occur, while `G' says that it
is generic. By generic we mean that one does not need to tune
initial values to find a trajectory in this class. In contrast,
`F' represents fine-tuned solutions, where one starts from a locus
of codimension at least one in the space of initial conditions.
Finally `C' stands for a solution which ``crosses over'' from one
fixed point to another.

\end{subsection}
\begin{subsection}{Curved internal space with flux}
We now turn to the three-dimensional phase spaces of models
with $k_1 = \pm 1$, which include, for suitable values of $c$,
flux compactifications on curved internal spaces. Although
the method remains the same, one faces new practical challenges
here. Three-dimensional phase spaces are harder to visualize,
especially in printed media which only allow static two-dimensional
projections. Therefore we have only included one ``three-dimensional''
plot, figure \ref{zwei},
for illustrational purposes, and summarize our
results in tables \ref{t.2} and \ref{t.4}.

For the classification of the trajectories according to their accelerational phases, it turns out that projecting the full solutions onto the $(x,y)$-plane is a very useful tool. This is due to the equations describing the acceleration region \refeq{1.12} being independent of $z$, so that the number of accelerating phases can directly be read off from this projection. In order to determine which regions of initial conditions may give rise to different behaviors of solutions it is further helpful to study the vector field generating the trajectories. To find the fine-tuned solutions, knowledge about the (Q)FP structure of the phase space is crucial.

\begin{figure}[t]
\renewcommand{\baselinestretch}{1}
\begin{center}
\leavevmode
\epsfxsize=0.55\textwidth \;
\epsffile{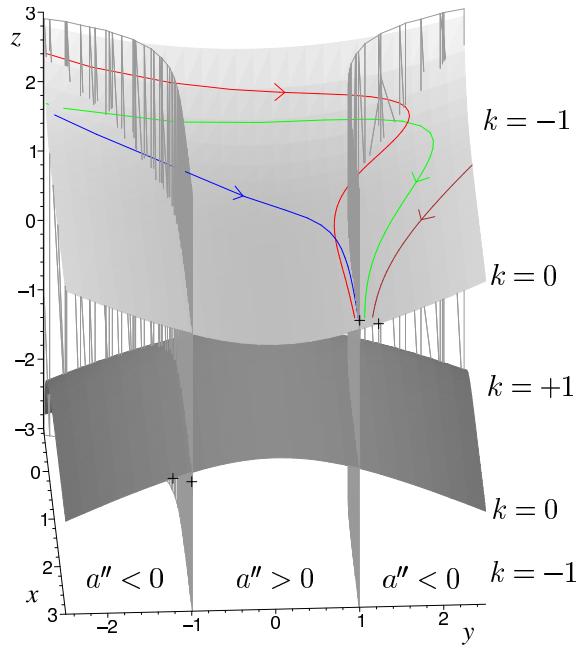}
%
\end{center}
\parbox[c]{\textwidth}{\caption{\label{zwei}{\footnotesize Phase portrait for hyperbolic flux compactifications of M-theory, $c = \frac{3}{\sqrt{7}}$ (and taking $\tilde{b} = 1$). The diagram shows one example for each class of trajectories encountered in the $k = -1$ region. From right to left: eternal deceleration, eternal deceleration preceeded by a transient accelerating phase, eternal acceleration and eternal acceleration with an earlier transient accelerating phase. }}}
\end{figure}

Let us illustrate the properties of cosmologies with a
three-dimensional phase space using the most interesting
examples, expanding open universes in hyperbolic flux compactifications
of M-theory, see figure \ref{zwei}. Here the horizontal and
vertical axis correspond to $y=\dot{\phi}$ and $z=\dot{\alpha}$
as before, while the new direction orthogonal to this surface
is the new phase space variable $x=\phi$. It is useful to compare
with the phase portraits of figure \ref{eins}, which correspond to the
$x\rightarrow + \infty$ limit of the three-dimensional plot.
As is apparent from this comparison, the two surfaces aligned to
the $(x,y)$-plane are the subspaces corresponding to flat universes.
Thus again we have disjoint regions populated by expanding open,
closed, and contracting open
universes. The two vertical surfaces enclose
the region of acceleration between them. This is the generalization of
the gray strip of figure \ref{eins}.
According to table \ref{t.1} two (pairs of) QFPs exist for
$k_1 = -1, 1 < c < \ft{2}{\sqrt{3}}$, namely QFP${}_4$ and QFP${}_5$.
These are actually located at $x=\phi=\infty$, but we have represented
their projections into the displayed region of the phase space by
little crosses. The QFP${}_4$ is a
real attractor in the open universe region, located on the
boundary between acceleration and deceleration, while the
QFP${}_5$ is a saddle point on the surface of flat universes,
located in the deceleration region. It is attractive for
flat universes, but repulsive normal to the $k=0$ surface.
Therefore expanding open universes approach the QFP${}_4$ with
eternal deceleration or acceleration, depending on from which side
they come.

In order to analyze whether periods of transient accelerated expansion
exist, one must plot and analyze the phase space flow. We found
four different types of generic behavior: eternal deceleration
(dark brown line), eternal deceleration with one period of
transient acceleration (green line), eternal acceleration (blue
line), eternal acceleration with one period of transient
acceleration (red line). The existence of generic M-theory
solutions with two periods of accelerated expansion (red line) is
a new feature which does not exist for a single exponential
potential. Therefore one needs both flux and a curved internal
space to have such behavior. We also note that figure \ref{zwei}
clearly illustrates that the concept of a QFP introduced in the
last section is indeed useful as the trajectories flow to $x
\rightarrow \infty$ at the values $[ \pd^*, \ad^* ]$ of the
QFP$_4$.

The classification of the possible trajectories in the other $c$
regimes and $k$ regions proceeds completely analogously.
Our results for hyperbolic ($k_1 = -1$) and spherical $(k_1 = +1)$
internal space with flux are summarized in tables \ref{t.2} and
\ref{t.4}, respectively.
An interesting feature with spherical compactifications is that
the $k=0$ surface becomes connected, so that the regions of
expanding and contracting open universes are not separated any
more. In this case there are flat and open universe solutions
which start in the $z>0$ region and have a period of transient
accelerated expansion, and then go down to the $z<0$ region, where
they display a period of transient decelerated
contraction.\footnote{Since our models satisfy the null energy
condition, flat and open cosmologies cannot switch from
contraction to expansion, but the opposite is allowed. Closed
cosmologies are not constrained.}
The FP$_3$ occurring for $k_1 = +1$ and $c > \sqrt{3}$ has a
special status as solutions that evolve at the FP correspond to
true de Sitter solutions. This behavior is denoted by the subscript
`$_{\rm dS}$' appearing in table \ref{t.4}. 
It is only available upon fine-tuning, as the
solution has to start at this FP. The two copies of FP$_3$, one in the $z<0$ region and 
one in the $z>0$ region, are connected by a special cross-over solution which runs through the 
$k=+1$ domain from the lower to the upper FP$_3$.
\begin{table}[tp!]
\renewcommand{\baselinestretch}{1}
\begin{tabular*}{\textwidth}{@{\extracolsep{\fill}} ccccccccccc} \hline \hline
    &    & \multicolumn{2}{c}{} & \multicolumn{2}{c}{acc.} & \multicolumn{2}{c}{dec.} & acc. & dec. &  \\
$c$     &  $k$   & \multicolumn{2}{c}{singular} & \multicolumn{2}{c}{expan.} & \multicolumn{2}{c}{expan.} & contr. & contr. & cyclic \\
        &        &   0   &   1  & 1 & 2 &  0 & 1 & 0 & 0 & $\infty$ \\ \hline
$0 < c < 1$
            & $-1$ &  G  & $-$ &  G  &  G  & $-$ & $-$ & $-$ & $-$ & $-$ \\
            & $ 0$ &  G  & $-$ &  G  &  G  & $-$ & $-$ & $-$ & $-$ & $-$ \\
            & $+1$ & G,F &  G  & G,C &  G  &  F  &  F  &  F  &  C  & $-$ \\ \hline
$ 1 < c \le \frac{2}{\sqrt{3}}$
            & $-1$ &  G  & $-$ &  G  &  G  & G,C &  G  &  C  & $-$ & $-$ \\
            & $ 0$ &  G  & $-$ & $-$ & $-$ &  G  &  G  & $-$ & $-$ & $-$ \\
            & $+1$ & G,F &  G  & $-$ & $-$ & $-$ & $-$ &  F  & $-$ & $-$ \\ \hline
$\frac{2}{\sqrt{3}} < c < \sqrt{3}$
            & $-1$ &  G  & $-$ & $-$ & $-$ & $-$ & $-$ &  C  & $-$ & G,C \\
            & $ 0$ &  G  & $-$ & $-$ & $-$ &  G  &  G  & $-$ & $-$ & $-$ \\
            & $+1$ & G,F &  G  & $-$ & $-$ & $-$ & $-$ &  F  & $-$ & $-$ \\ \hline
$\sqrt{3} < c < \frac{3}{2} \sqrt{3}$
            & $-1$ &  G  & $-$ & $-$ & $-$ & $-$ & $-$ &  C  & $-$ & G,C \\
            & $ 0$ &  G  & $-$ & $-$ & $-$ &  G  &  G  & $-$ & $-$ & $-$ \\
            & $+1$ & G,F &  G  & $-$ & $-$ & $-$ & $-$ &  F  & $-$ & $-$ \\ \hline
$\frac{3}{2} \sqrt{3} \le c < 3$
            & $-1$ &  G  & $-$ &  G  &  G  & G,C &  G  &  C  & $-$ & $-$ \\
            & $ 0$ &  G  & $-$ & $-$ & $-$ &  G  &  G  & $-$ & $-$ & $-$ \\
            & $+1$ & G,F &  G  & $-$ & $-$ & $-$ & $-$ &  F  & $-$ & $-$ \\ \hline
$ 3 < c $
            & $-1$ &  G  & $-$ &  G  &  G  & $-$ & $-$ & $-$ & $-$ & $-$ \\
            & $ 0$ &  G  & $-$ &  G  &  G  & $-$ & $-$ & $-$ & $-$ & $-$ \\
            & $+1$ & G,F &  G  & G,C &  G  &  F  &  F  &  F  &  C  & $-$ \\
\hline \hline
\end{tabular*}
\renewcommand{\baselinestretch}{1}
\parbox[c]{\textwidth}{\caption{\label{t.2}{\footnotesize Classification of possible cosmological solutions for $k_1 = -1$ according to their late-time behavior and their number of periods of accelerated expansion.
}}}
\end{table}
\begin{table}[tp!]
\renewcommand{\baselinestretch}{1}
\begin{tabular*}{\textwidth}{@{\extracolsep{\fill}} ccccccccccc} \hline \hline
    &    & \multicolumn{2}{c}{} & \multicolumn{2}{c}{acc.} & \multicolumn{2}{c}{dec.} & acc. & dec. &  \\
$c$     &  $k$   & \multicolumn{2}{c}{singular} & \multicolumn{2}{c}{expan.} & \multicolumn{2}{c}{expan.} & contr. & contr. & cyclic \\
        &        &   0   &   1  & 1 & 2 &  0 & 1 & 0 & 0 & $\infty$ \\ \hline
$ 0 < c \le \sqrt{3}$
            & $-1$ &  G  &  G  & $-$ & $-$ & $-$ & $-$ & $-$ & $-$ & $-$ \\
            & $0$  &  G  &  G  & $-$ & $-$ & $-$ & $-$ & $-$ & $-$ & $-$ \\
            & $+1$ &  G  &  G  & $-$ & $-$ & $-$ & $-$ & $-$ & $-$ & $-$ \\ \hline
$ \sqrt{3} < c < \frac{3}{2} \sqrt{3}$
            & $-1$ & G,F &  G  &  F  & $-$ & $-$ & $-$ &  C  & $-$ & G,C \\
            & $ 0$ &  G  &  G  &${\rm F}_{\rm dS}$& $-$ &  G  & G,C & $-$ &${\rm C}_{\rm dS}$& $-$ \\
            & $+1$ & G,F &  G  &  C  & $-$ & $-$ & $-$ &  F  & $-$ & $-$ \\ \hline
$\frac{3}{2} \sqrt{3} \le c < 3 $
            & $-1$ & G,F &  G  & G,F & $-$ & G,C &  G  &  C  & $-$ & $-$ \\
            & $ 0$ &  G  &  G  &${\rm F}_{\rm dS}$& $-$ &  G  & G,C & $-$ &${\rm C}_{\rm dS}$& $-$ \\
            & $+1$ & G,F &  G  &  C  & $-$ & $-$ & $-$ &  F  & $-$ & $-$ \\ \hline
$ 3 < c $
            & $-1$ & G,F &  G  & G,F & $-$ & $-$ & $-$ & $-$ & $-$ & $-$ \\
            & $ 0$ &  G  &  G  &G,${\rm F}_{\rm dS}$& $-$ & $-$ & $-$ & $-$ &${\rm C}_{\rm dS}$& $-$ \\
            & $+1$ & G,F &  G  & G,C & $-$ &  F  &  F  &  F  &  C  & $-$ \\
\hline \hline
\end{tabular*}
\renewcommand{\baselinestretch}{1}
\parbox[c]{\textwidth}{\caption{\label{t.4}{\footnotesize Classification of possible cosmological solutions for $k_1 = +1$ according to their late-time behavior and their number of periods of
accelerated expansion.
The subscript `$_{\rm dS}$' indicates the existence of true de Sitter solutions if we fine-tune our initial conditions to be at the FP$_{3}$.
In addition there are also fine-tuned and cross-over solutions which approach this fixed point at late times. }}}
\end{table}

Finally let us make a remark about this classification. One
should be aware of the fact that our results depend on a mixture
of analytical and numerical results.
Knowledge about the (quasi) fixed points basically determines what kind of
late-time behavior is possible -- the solutions either run to a
(Q)FP or the boundary of the phase space. Numerical investigation
is needed mainly to determine how many periods of accelerated
expansion can occur.
Consistency checks\footnote{One example of an internal consistency
check is to compute $k$ as a function of the fields. Analytically
we know that $k$ takes the discrete values $0,1,-1$, and when
plugging in numerical solutions deviations should be small.}
indicate that the numerical results are trustworthy if we evolve
trajectories with initial conditions in the $z > 0$ region forward
in time.
However, when following trajectories back in time
we find indications that the numerics becomes untrustworthy below a certain critical time $t_c$ where the numerical value of $t_c$ depends on the initial conditions chosen. Due to this fact it is hard to make definite statements about the early-time behavior of our trajectories.

%
%
%
%
\end{subsection}
\end{section}
\begin{section}{Physics at the Fixed Points}

In this section we use the information about the (Q)FPs obtained in section 2.4 to investigate the properties of
solutions asymptotically approaching these points.

\subsubsection*{Fixed point solutions in cosmological time}

In order to find the
physical interpretation of our solutions we need to go back from the time coordinate
$t$, which was introduced to simplify the analysis of the equations
of motion, to the cosmological time $\tau$.
Our starting point is eq. \refeq{1.17a}. At a (Q)FP, this equation is easily integrated and yields the following solutions parametrized by the lapse function time $t$:
\be\label{4.1}
\begin{array}{lll}
\alpha(t) = \ad^* \, t + \zeta \, , \quad &  \quad \phi(t) = \pd^* \, t \, ,  \quad & \mbox{for the QFPs} \, , \\
\alpha(t) = \ad^* \, t \, , \quad & \quad \phi(t) = \phi^*  \, , \quad & \mbox{for FP$_{3}$} \, .
\end{array}
\ee
This is the most general solution of eq. \refeq{1.17a} as in the first case the constant of integration appearing in the $\phi(t)$ equation can always be absorbed by a shift of the time variable $t$. In the case of ${\rm FP}_3$ we have used this freedom to absorb the constant appearing in the $\alpha(t)$ equation, since  $\phi^*$ cannot be set to zero. Concerning the remaining constant of integration, $\zeta$, we have to distinguish between the cases where the QFP is located on the $k = 0$ surface or in the $k = \pm 1$ region. In the former case $\zeta$ is indeed an undetermined constant, while in the latter case it is fixed by Friedmann's equation:
\be\label{4.1a}
e^{2 \zeta} = \frac{2 \, k \, c^2}{\bt^2 \left( c^2 - 9 \right) }
\quad \mbox{for QFP${}_{1,6}$}
\,  , \quad \quad \quad
e^{2 \zeta} = \frac{k}{1 - c^2}
\quad \mbox{for QFP${}_4$} \, .
\ee

Switching to
the cosmological time $\tau$ via $\rmd \tau = N(t) \rmd t$ with $N(t)$ being the lapse function given in eq. \refeq{1.9}, we can use the explicit expression for $\phi(t)$ at the (Q)FP to calculate the relation between the lapse function time $t$ and the cosmological time $\tau$:
\be\label{4.3}
\begin{array}{rclll}
t & = & \frac{c}{3 \pd^*} \, \ln \left( \frac{3 \pd^* \bt}{\sqrt{2} c} \, \tau \right) \quad & \mbox{for QFP$_{1,2,6,7}$} \; , \\
t & = & \frac{1}{c \pd^*} \, \ln \left( c \, \pd^* \tau \right) & \mbox{for QFP$_{4,5}$ } \; , \\
t & = & \frac{\bt}{\sqrt{2}} \,  \tau & \mbox{for FP$_{3}$ } \, .
\end{array}
\ee
%
%
Using these relations it is now straightforward to write down the cosmological solutions at the (Q)FPs parametrized in cosmological time. Going back to the scale factor $a(\tau) = e^{\alpha(t(\tau))}$ we find \\
\bea
\label{4.4}
{\rm QFP}_{1,2,6,7}: \quad
&
a(\tau)  =  \left( \frac{3 \pd^* \bt}{\sqrt{2} c} \right)^{\frac{\ad^* c }{3 \pd^*}} \, \tau^{\frac{\ad^* \, c }{3 \, \pd^*}} \, e^{\zeta} \, , \quad
&
\phi(\tau)  =  \frac{c}{3} \, \ln \left( \ft{3 \pd^* \, \bt}{\sqrt{2} \, c} \, \tau \right) , \\
 & & \nonumber\\
\label{4.5}
{\rm QFP}_{4,5}: \quad
&
a(\tau) = \left( c \, \pd^* \right)^{\frac{\ad^*}{c \pd^*}} \, \tau^{\frac{\ad^*}{c \pd^*}} \, e^{\zeta} \, , \quad
&
\phi(\tau) = \frac{1}{c} \ln \left( c \,  \pd^* \, \tau \right) , \\
 & & \nonumber\\
\label{4.6}
{\rm FP}_{3}: \quad
&
a(\tau) = \exp \left( \frac{\bt \, \ad^*}{\sqrt{2}}  \, \tau \right) \, , \quad \quad \quad
&
\phi(\tau) = \phi^* \, .
\eea
%
%
Note that the evolution of the metric and the scalar fields as
functions of $\tau$ is well behaved and universal for $\tau > 0$.
This confirms that our choice of lapse function and the notion
of a QFP (which depends on the choice of lapse function)
is physically meaningful. Switching between $t$ and $\tau$ does not affect
the qualitative late-time features of the solutions: they are either
governed by the (Q)FP which in cosmological time corresponds to the asymptotic behavior given above, or end up in a big crunch singularity
which is a coordinate-independent notion. This assures that the
classification of solutions with respect to $t$ carries over
to the physically relevant cosmological time $\tau$.

\subsubsection*{Properties of the fixed point solutions}
All QFP solutions show a run-away behavior of the scalar field, $\phi \sim \ln \tau$, which
slows down to a stop as $\tau \rightarrow \infty$.
The scale factor follows a power law expansion $a \sim \tau^\Delta$,
where the ``critical exponents'' $\Delta$ depend on the model
parameter $c$ only. The values of $\Delta$ at the (Q)FPs are summarized in the column `$a$' of table \ref{t.5}.
Accelerating expansion corresponds to $\Delta > 1$
and hence requires $c>3$ for QFP${}_{2,7}$ or $c<1$ for QFP${}_{5}$. These, however,
are outside the range of compactifications. From compactifications we may obtain
at most $\Delta=1$. This corresponds to
a QFP on the boundary of the region of acceleration
and deceleration. In the previous section
we found that such a QFP can
be approached from the acceleration region. In this case the acceleration
is eternal but approaches zero so fast that one does not
obtain a significant growth of the scale factor. This was also observed
in \cite{Chen:2003ij,Ish1}.

Besides the QFPs we have also
the de Sitter fixed point FP${}_3$, which does not exist for
potentials with one exponential term. Here the scale factor
grows exponentially, $a \sim \exp \tau$, while the scalar
field has a finite value $\phi^*$. However,
this FP does not exist for compactifications: $\phi^*$
goes to zero for $c\rightarrow \infty$ and moves off to infinity when
approaching the maximal value for compactifications $c=\sqrt{3}$ from above.
Moreover, it is a saddle point. Hence eternal expansion
requires an infinite fine-tuning,  which means to pick exact initial
values. Generating any significant
growth of the scale factor by passing near the saddle point still
requires a drastic fine-tuning.\footnote{A numerical solution placed directly on FP$_3$ with 10 digits precision triggers 10 e-folds of expansion before leaving the FP due to accumulated numerical errors.}

We can also compute the density parameter $\Omega$ and the equation of state
parameter $w$ for the (Q)FP solutions.
Recall the standard expressions for the density and pressure
of a scalar field,
\be\label{4.7}
\rho =  \,  \phi^{{\prime}^2} + 2 \, V(\phi) \,,\quad
p =  \,  \phi^{{\prime}^2} - 2 \, V(\phi)  \, ,
\ee
where the prime denotes the derivative with respect to cosmological time
$\tau$. Defining the Hubble parameter $H := \frac{a^\prime}{a}$ and introducing
the density parameter $\Omega := \frac{\rho}{3 H^2}$ Friedmann's
equation \refeq{1.6} can be written as
\be\label{4.8}
\Omega = 1 + \frac{k}{a^{{\prime}^{2}}} \, .
\ee
Here $\Omega = 1$ is the critical density corresponding to a flat universe,
while $\Omega > 1 $ and $\Omega < 1$ correspond to a closed and open universe,
respectively. Taking the $k$ values for the (Q)FPs given in
table \ref{t.1} together with the solutions for $a(\tau)$ obtained by
evaluating the eqs. \refeq{4.4}, \refeq{4.5} and \refeq{4.6}, we can use this relation to find the density
parameter $\Omega$ at the (Q)FP. The results depend on the parameter
$c$ only and are summarized in the column denoted `$\Omega$' of table \ref{t.5}.

Furthermore, the density and pressure of the scalar field are related by an equation of
state, $p = w \rho$, where $p$ and $\rho$ are given in (\ref{4.7}).
%
%
At a (Q)FP $w$ takes a constant value, which is an
intrinsic property of the (Q)FP,
in the sense that it depends on the model parameter $c$ only. The
values are again listed in table \ref{t.5}. In the compactification range $ 1 \le c \le \sqrt{3}$, we typically obtain $\Delta \le 1$, $w \ge - \frac{1}{3}$ and $\Omega = {\cal O}(1)$.
There is a close relation between the
stability properties of the (Q)FP and its $w$ value.
While $w < - \frac{1}{3}$ always corresponds
to an attractor, $w > - \frac{1}{3}$ characterizes a saddle point. For the boundary
value $w = - \frac{1}{3}$ the (Q)FP is attractive when located in the $k = -1$
region (open universe) while it is a saddle point if $ k = +1$
(closed universe). An $\Omega = 1$ attractor sitting at the $k=0$ surface requires $w < - \frac{1}{3}$, $\Delta > 1$ and is only realized for $k_1 = -1$, $ c < 1$ and $ c > 3$.
\begin{table}[tp!]
\begin{tabular*}{\textwidth}{@{\extracolsep{\fill}} cccccccc } \hline \hline
(Q)FP & $k_1$ & $c$ & type & $a$ & $\Omega$ & $w$ &  $r_{\rm
 int}$  \\ \hline
1  & $0$ & $0<c\le\sqrt{3}$ & attractor & $\tau$ & $\frac{c^2}{9}$ & $- \,
 \frac{1}{3}$ & $\tau^{\frac{2}{3 m}}$ \\
   &         & $\sqrt{3}<c<3$   & attractor & $\tau$ & $\frac{c^2}{9}$ & $- \,
 \frac{1}{3}$ & $-$ \\
   &         & $3<c$            & saddle    & $\tau$ & $\frac{c^2}{9}$ & $- \,
 \frac{1}{3}$ & $-$ \\ \hline
2  & $0 $ & $0<c\le\sqrt{3}$ & $-$       & $-$             & $-$ & $-$
            & $-$ \\
   &          & $\sqrt{3}<c<3$   & saddle    & $\tau^{\frac{c^2}{9}}$ & $1$ &
 $-1+\frac{6}{c^2} $ & $-$ \\
   &          & $3<c$            & attractor & $\tau^{\frac{c^2}{9}}$ & $1$ &
 $-1+\frac{6}{c^2} $ & $-$ \\ \hline
3  & $+1$ & $0<c\le\sqrt{3}$ & $-$    & $-$        & $-$ & $-$  & $-$ \\
   &          & $\sqrt{3}<c$     & saddle &
 $e^{\frac{\tilde{b}\sqrt{c^2-3}}{c\sqrt{3}}\tau}$ & $1$ & $-1$ & $-$ \\ \hline
4  & $-1$ & $0<c<1$          & saddle    & $\tau$ & $\frac{1}{c^2}$ & $- \,
 \frac{1}{3}$ & $-$  \\
   &          & $1<c<\sqrt{3}$   & attractor & $\tau$ & $\frac{1}{c^2}$ & $- \,
 \frac{1}{3}$ & $\tau^{\frac{2}{m c^2}}$ \\
   &          & $\sqrt{3}\le c$  & $-$       & $-$    & $-$             & $-$
             & $-$ \\ \hline
5  & $-1$ & $0<c<1$         & attractor & $\tau^{\frac{1}{c^2}}$ & $1$ & $-1
 +\frac{2 c^2}{3}$  & $-$ \\
   &          & $1<c<\sqrt{3}$  & saddle    & $\tau^{\frac{1}{c^2}}$ & $1$ & $-1
 +\frac{2 c^2}{3}$  & $\tau^{\frac{2}{m c^2}}$ \\
   &          & $\sqrt{3}\le c$ & $-$       & $-$             & $-$ & $-$
             & $-$ \\ \hline
6  & $\pm 1$ & $0<c\le\sqrt{3}$ & $-$       & $-$ & $-$             & $-$
             & $-$ \\
   &             & $\sqrt{3}<c<3$   & attractor & $\tau$ & $\frac{c^2}{9}$ & $
 -\, \frac{1}{3}$ & $-$ \\
   &             & $3<c$            & saddle    & $\tau$ & $\frac{c^2}{9}$ & $
 -\, \frac{1}{3}$ & $-$ \\ \hline
7  & $\pm 1$ & $0<c\le\sqrt{3}$ & $-$       & $-$             & $-$ & $-$
                & $-$  \\
   &              & $\sqrt{3}<c<3$   & saddle    & $\tau^{\frac{c^2}{9}}$ & $1$
 & $-1+\frac{6}{c^2}$ & $-$ \\
   &              & $3<c$            & attractor & $\tau^{\frac{c^2}{9}}$ & $1$
 & $-1+\frac{6}{c^2}$ & $-$ \\ \hline \hline
\end{tabular*}
\renewcommand{\baselinestretch}{1}
\parbox[c]{\textwidth}{\caption{\label{t.5}{\footnotesize Summary of the properties of cosmological solutions parametrized in cosmological time $\tau$ at the (quasi) fixed points found in  section 2.4.}}}
\end{table}

\subsubsection*{Decompactification of the extra dimensions}
Except at the saddle point FP${}_3$, the scalar field $\phi$ cannot
be stabilized and exhibits run-away behavior. For the values
$1 \le c \le \sqrt{3}$, which correspond to compactifications
of higher-dimensional theories, this implies the decompactification
of the internal manifold. In order to see whether such behavior is
in immediate conflict with observations, we need to characterize this
quantitatively. From eqs. (\ref{Metric1}) -
(\ref{Metric3}) we obtain
\be\label{4.11}
V_{\rm int} = \int \rmd^m y \, \sqrt{-\hat{g}} = e^{m \, \beta(t)} \int \rmd^m y \sqrt{-\hat{\omega}} = e^{m \beta(t)} \, ,
\ee
which shows that the radius of the internal manifold behaves as
$r_{\rm int} \sim e^{\beta(t)}$.
Taking the definition of $\beta(t)$ given in eq. \refeq{1.2a} and substituting
the solutions \refeq{4.4} and \refeq{4.5} we find a polynomial increase of
$r_{\rm int} \sim \tau^q$, where $q < 1$ for all cases. The explicit
values of $q$ are given in the column `$r_{\rm int}$' of table \ref{t.5}. The
result $q < 1$ implies that the expansion of the internal space
always decelerates.

\subsubsection*{M-theory example}

As an explicit example let us consider the hyperbolic flux
compactification of M-theory  ($m = 7, c = \frac{3}{\sqrt{7}}$).
From table \ref{t.5} we find that there exist
solutions attracted to QFP${}_4$. At this QFP the universe
expands at a constant rate $a \sim \tau$, while the radius of the internal manifold
exhibits decelerated growth, $r_{\rm int} \sim \tau^{\frac{2}{9}}$. The energy density
and equation of state parameter have the QFP values
$\Omega = \ft79 \approx .77$ and $w=-\ft13$, respectively.
\end{section}
\begin{section}{Discussion and Conclusions}
In this paper we used phase space methods to investigate the properties of Friedmann-Robertson-Walker cosmological solutions with a scalar field governed by a potential with two exponential terms. The form of the potential is motivated by the compactification of M-theory on a maximally symmetric internal manifold including flux. Our main result is the complete classification of all possible open, flat, and closed cosmologies according to their late-time behavior and the number of periods of accelerated expansion. The concept of a quasi fixed point introduced in section 2.4 proves very useful for this purpose as it captures the asymptotic behavior of run-away solutions. This allowed us to calculate numerical values for cosmological observables at these (quasi) fixed points (given in table \ref{t.5}).  We also discovered a new class of open universe solutions occurring in hyperbolic flux compactifications of M-theory. These undergo two periods of accelerated expansion where the first is transient while the second is eternal and asymptotes to zero.
The quasi fixed point governing these solutions resides very close to the hypersurface of flat universes. Therefore all solutions approaching it are rather flat at late times.

The main motivation of our investigation is to describe the current universe by interpreting the scalar field as a quintessence. As already noted by many authors before, this model has the necessary qualitative feature that it generically induces accelerated expansion. The acceleration can be transient, eternal or eternally cyclic switching between acceleration and deceleration.

Comparing the (quasi) fixed point solutions of our model with cosmological measurements, however, needs some caution. Since the accelerated expansion is a relatively recent phenomenon in the history of the universe, it is clear that we cannot be arbitrarily close to a (quasi) fixed point yet. We expect that even in the case where our universe approaches such a point, the cosmological parameters computed in section 4 will receive corrections. Therefore it is hard to draw conclusions by comparing these quantities with observable data, as the (quasi) fixed points might only rule the universe some time in the future.

A crucial ingredient missing in our setup is the component of (dark and ordinary) matter. Ref. \cite{GKL,KM-RST} list some options how this sector could be obtained from string theory, but it still remains a big challenge to derive the appropriate dark energy and matter from a fundamental theory simultaneously. The inclusion of matter probably also modifies some of the quantities calculated at the (quasi) fixed points, although the qualitative features should remain the same.

As investigated in ref. \cite{GKL}, models coming from hyperbolic flux compactifications of M-theory with matter added by hand allow fine-tuned solutions, which are  (marginally) compatible with cosmological data.
Let us point out, that
the double exponential potential derived from hyperbolic flux compactifications satisfies the tracker condition $\Gamma = \frac{V_{, \phi \phi} V}{(V_{, \phi})^2} >1$ \cite{tracker}. For the potential (\ref{1.2}) with $k_1 = -1$, $\Gamma > 1$ is always satisfied. It deviates significantly from unity as long as both exponential terms are of comparable size and approaches $\Gamma = 1$ from above, once one of the exponential terms dominates. This raises the hope that this model could give rise to a tracker behavior, which means that for a wide range of initial conditions the quintessence energy density
approaches and follows the background (radiation or matter) density before
overtaking it and inducing a period of accelerated expansion.

Another crucial point in the models with a higher-dimensional origin is the problem of a spontaneous decompactification of the extra dimensions.
For the M-theory example of section 4, taking the age of the universe to be approximately $1.5 \times 10^{10}$ years,
and assuming that the
solution describes the whole evolution except the big bang and primordial inflation,
we find that the
radius of the extra dimension should have increased by about a factor $10^4$.
The current observation of three large dimensions is then
compatible with a start value for $r_{\mscr{int}}^{-1}$
well below the Planck scale, even if one uses the naive estimate
$m_{KK} \sim r_{\mscr{int}}^{-1}$ for the masses of Kaluza-Klein
states. In the case of a hyperbolic internal manifold this should be a safe assumption, as these masses
seem to be much higher than suggested by this naive estimate \cite{KM-RST}.

Our system also allows for an interpretation in terms of the very early universe.\footnote{In fact, for the work of ref. \cite{Halliwell} which
we generalize, this was the main motivation.} In this case matter effects are usually neglected. To realize a power law inflation scenario and solve the horizon and flatness
problems without fine-tuning, one needs
an attractive (quasi) fixed point with $\Delta >1$ on the $k=0$ surface \cite{Halliwell}. Comparing to table \ref{t.5} we find that there are such points, {\it viz.} QFP${}_2$ and QFP${}_7$ for $c>3$
and QFP${}_5$ for $0<c<1$. Solutions attracted to them are
driven to $\Omega = 1$, even when $k=\pm 1$.
The de Sitter fixed point, FP${}_3$, allows exponential expansion,
but it is a saddle point and therefore
the solutions need a drastic fine-tuning to produce inflation.
Unfortunately none of those fixed points occur for scalar potentials
obtainable from compactifications on maximally symmetric spaces with flux.

While ten- and eleven-dimensional supergravities satisfy
the strong energy condition, $\rho + 3 p \geq 0$, which
forbids cosmic acceleration, an action of the form
\refeq{1.1} only needs to satisfy the null energy condition,
$\rho + p \geq 0$, which leaves some room for acceleration.
However, if a theory is obtained by dimensional reduction
of a theory obeying the strong energy condition,
the lower-dimensional theory cannot have
de Sitter vacua \cite{GMN}.
This could be misinterpreted as a
no-go theorem for accelerated expansion, but, as pointed
out in  ref. \cite{Townsend:2003fx}, acceleration is
nevertheless possible, if the internal manifold is
time-dependent. However, in all known cases the acceleration
is either transient, or eternal but going to zero asymptotically.
In our tables (for compactifications, $1 \leq c \leq \sqrt{3}$) this is reflected by the values
of $w$ for (quasi) fixed point solutions, which
always satisfy and often saturate the bound $w \geq - \ft13$
set by the strong energy condition. This indicates how the
strong energy condition satisfied by the higher-dimensional theory
restricts the amount of acceleration obtainable in time-dependent
compactifications, by requiring it to converge to zero for
late times. Note that this fits nicely with the fact that
the internal space always decompactifies in this limit.\footnote{This observation agrees with the more general analysis \cite{Giddings} which argued that compactifications leading to an accelerated expansion of the 4-dimensional space-time either decompactify at late-times or end in a big crunch singularity.} Therefore it is natural that the asymptotic behavior can be understood
in terms of the higher-dimensional theory.

Let us finally discuss the main approaches for avoiding the consequences of the strong energy condition valid in the higher-dimensional theory and
for stabilizing moduli. Constructions involving non-perturbative effects
of string/M-theory, like branes, orientifolds, and instantons
can produce inflation and stabilize the moduli \cite{KKLT},
but seem to require a significant discrete fine tuning. Therefore it remains
important to investigate various other directions.
For example, the dimensional reduction
on group manifolds also leads to multi exponential potentials allowing
accelerating cosmologies \cite{Rome}.
One promising strategy to avoid the strict
limits on the amount of acceleration obtainable in compactifications
is to consider generalized compactifications, where one twists
the theory by a symmetry of the equations of motion which is not
a symmetry of the action. This leads to lower-dimensional theories
which do not
have an action \cite{Bianchi}.\footnote{Supergravity theories
without an action have also been constructed without reference
to dimensional reduction \cite{SC25}.}
It would be interesting to explore whether such constructions
can lead to
double exponential potentials with $c$ being outside the
range $1 \leq c \leq \sqrt{3}$ valid for maximally symmetric
spaces. Another interesting direction was pointed out
by the authors of refs. \cite{BAN}, who argue that the `central region'
in the moduli space, which cannot be mapped
to a perturbative regime by duality transformations, might
give viable cosmologies without fine-tuning. One particular approach for exploring this region is the inclusion
of states which become light at special points in the moduli
space \cite{us1}. The effects of such extra states improve on
the moduli problem, seem to be important for
vacuum selection, and help to get cosmic acceleration
\cite{us2,KLLMMS}.

\end{section}
\subsection*{Acknowledgments}
We would like to  thank
I.P. Neupane for an exchange of e-mails which motivated us to undertake this investigation. This work is supported by
the DFG within the `Schwerpunktprogramm Stringtheorie'.
F.S. acknowledges a scholarship from the
`Studienstiftung des deutschen Volkes'. L.J. was also supported by the Estonian Science Foundation Grant No 5026.
\end{document}